\documentclass[reprint,amsmath,amssymb,aps,prb,superscriptaddress]{revtex4-1}
\pdfoutput=1
\usepackage{array}
\usepackage{bm}
\usepackage{color}
\usepackage{graphicx}
\usepackage{hyperref}
\hypersetup{colorlinks=true,allcolors=blue}
\usepackage{natbib}
\usepackage{newtxtext}
\usepackage{newtxmath}
\usepackage[caption=false]{subfig}
\begin{document}

\title{Magnetic Order with Fractionalized Excitations: Applications to $\mathrm{Yb}_2 \mathrm{Ti}_2 \mathrm{O}_7$}

\author{Li Ern Chern}
\affiliation{Department of Physics, University of Toronto, Toronto, Ontario M5S 1A7, Canada}

\author{Yong Baek Kim}
\affiliation{Department of Physics, University of Toronto, Toronto, Ontario M5S 1A7, Canada}
\affiliation{Canadian Institute for Advanced Research/Quantum Materials Program, Toronto, Ontario M5G 1Z8, Canada}
\affiliation{School of Physics, Korea Institute for Advanced Study, Seoul 130-722, Korea}

\begin{abstract}
A recent inelastic neutron scattering experiment on $\mathrm{Yb}_2 \mathrm{Ti}_2 \mathrm{O}_7$ uncovers an unusual scattering continuum in the spin excitation spectrum despite the splayed ferromagnetic order in the ground state. While there exist well defined spin wave excitations at high magnetic fields, the one magnon modes and the two magnon continuum start to strongly overlap upon decreasing the field, and eventually they become the scattering continuum at zero field. Motivated by these observations, we investigate the possible emergence of a magnetically ordered ground state with fractionalized excitations in the spin model with the exchange parameters determined from two previous experiments. Using the fermionic parton mean field theory, we show that the magnetically ordered state with fractionalized excitations can arise as a stable mean field ground state in the presence of sufficiently strong quantum fluctuations. The spin excitation spectrum in such a ground state is computed and shown to have the scattering continuum. Upon increasing the magnetic field, the fractionalized magnetically ordered state is suppressed, and is eventually replaced by the conventional magnetically ordered phase at high fields, which is consistent with the experimental data. We discuss further implications of these results to the experiments and possible improvements on the theoretical analysis.
\end{abstract}

\pacs{}

\maketitle
%\tableofcontents

\section{\label{introduction}Introduction}
The family of rare earth pyrochlore compounds is the exemplar of three dimensional frustrated magnets that offer tremendous opportunities for the discovery of exotic phases of matter. For instance, one of the most celebrated emergent phenomena in condensed matter physics is the identification of low energy excitations as effective magnetic monopoles\cite{nature06433,Gingras375,Morris411,Fennell415} in the classical spin ice materials $\mathrm{Ho}_2\mathrm{Ti}_2\mathrm{O}_7$ and $\mathrm{Dy}_2\mathrm{Ti}_2\mathrm{O}_7$, where the rare earth ion carries large angular momentum $J=8$ subjected to strong local Ising anisotropy. Many other pyrochlore compounds, such as $\mathrm{Yb}_2\mathrm{Ti}_2\mathrm{O}_7$, $\mathrm{Yb}_2\mathrm{Sn}_2\mathrm{O}_7$, $\mathrm{Tb}_2\mathrm{Ti}_2\mathrm{O}_7$, and $\mathrm{Pr}_2\mathrm{Zr}_2\mathrm{O}_7$,\cite{0034-4885-77-5-056501} to name a few, are characterized by strong quantum fluctuations and complex exchange interactions. They are less understood and currently still under intense experimental and theoretical investigations. Among the exciting prospects is the realization of the long-sought-after quantum spin liquid state,\cite{PhysRevB.69.064404,PhysRevLett.100.047208,PhysRevLett.108.037202,PhysRevB.86.104412,0034-4885-77-5-056501,1802.09198} which is devoid of magnetic order down to very low temperatures while exhibiting long range entanglement and fractionalized excitations, in these materials.

In $\mathrm{Yb}_2\mathrm{Ti}_2\mathrm{O}_7$, the low energy degrees of freedom of each $\mathrm{Yb}^{3+}$ ion is described by a Kramers doublet well separated from the first excited crystal field states,\cite{0953-8984-13-41-318,1742-6596-320-1-012065} so that the system can be treated as a pyrochlore array of pseudospin-$1/2$ moments (which are simply referred to as spins from now on). A number of experiments\cite{JPSJ.72.3014,ncomms1989,PhysRevB.93.064406,0953-8984-28-42-426002,PhysRevB.96.214415} have identified the splayed/noncollinear ferromagnetic order, where a net magnetization develops through canted spins, as the ground state of $\mathrm{Yb}_2\mathrm{Ti}_2\mathrm{O}_7$. The transition temperature is $\sim 0.2 \mathrm{K}$, which is about one order of magnitude less than the energy scale of the greatest exchange interaction. (It should be noted that there are other experiments\cite{PhysRevLett.88.077204,PhysRevB.70.180404,PhysRevLett.103.227202,PhysRevB.88.134428} that report a disordered ground state, but perhaps due to imperfection of the samples.) Yet a recent inelastic neutron scattering experiment\cite{PhysRevLett.119.057203} on $\mathrm{Yb}_2\mathrm{Ti}_2\mathrm{O}_7$ revealed some remarkably unconventional features in the magnetic ground state. While sharp one magnon modes and a two magnon continuum are well separated at high magnetic fields, they overlap with each other upon lowering the field, which leads to strong renormalization of the spin wave dispersions. As the field approaches zero, well defined spin wave dispersions can no longer be observed over a large region in the Brillouin zone, whereas a broad scattering continuum appears. This is interpreted in Ref.~\onlinecite{PhysRevLett.119.057203} as a consequence of one magnon decaying into two magnons, and their interaction is so strong that the linear spin wave theory breaks down.

The breakdown of magnons or spin wave excitations suggests the presence of strong quantum fluctuations despite the magnetic order in the ground state. Clearly, the semiclassical description of the ground state and the elementary excitations is not adequate for this system. Given that the scattering continuum seen in the experiment is reminiscent of the two spinon continuum in a quantum spin liquid, it may be useful to start from the extreme quantum limit or the spinon/parton representation of the spin exchange interactions. Such a description allows us to start from a quantum spin liquid phase with a built in two spinon continuum. In this spinon basis, the magnetically ordered state is obtained via confinement of spinons in the underlying spin liquid state. If the magnetically ordered state is at the verge of making a phase transition to a nearby spin liquid state, the confinement energy scale may be very small. It is then conceivable that the two spinon continuum could be seen above the small confinement energy scale, providing an alternative description of the scattering continuum seen in the experiment. The main difficulty with this approach, however, is that currently there is no well defined theoretical formulation to describe or compute the excitation spectrum of such ``loosely'' confined spinons as it is inherently a phenomenon in the strong coupling limit.

In this work, with the picture described above in mind, we investigate the possibility of a quantum spin liquid coexisting with a magnetic order, where the ground state is magnetically ordered, but the deconfined spinons exist as elementary excitations. Such a phase is possible in three dimensions while there could be a transition from the coexisting phase with deconfined spinons to a confined phase with conventional magnetic order upon changing the parameters of the model. In practice, the excitation spectrum of such a coexisting or fractionalized magnetically ordered phase would look similar to that of the magnetically ordered state with a small (spinon) confinement energy scale. Thus, if we take a more conservative stance, the coexisting phase may also be considered as a good approximate mean field description of the magnetically ordered state with a small confinement energy scale.

We consider the slave particle or parton mean field theory\cite{PhysRevB.65.165113,PhysRevB.83.224413,PhysRevB.80.064410,PhysRevB.95.054404} of the spin model with the exchange parameters obtained from the experimental data. According to these parameters, $\mathrm{Yb}_2\mathrm{Ti}_2\mathrm{O}_7$ is close to the classical phase boundary between the splayed ferromagnet and an antiferromagnetic state (see Fig.~\ref{classicalphasediagramfigure}, where the parametrizations of the spin exchange interactions from two different experiments,\cite{PhysRevX.1.021002,PhysRevLett.119.057203} dubbed Gaulin and Coldea parametrizations, are shown). We envision that a new quantum ground state such as the coexisting or a pure spin liquid state may emerge near the classical phase boundary. We examine the conditions under which the fractionalized magnetically ordered phase emerges as a stable mean field ground state and find that, as discussed below, it appears only when quantum fluctuations are sufficiently strong. A theoretical advantage of considering such a coexisting phase is that we can compute the excitation spectrum at the mean field level.

For this purpose, we first notice that the spin Hamiltonian of $\mathrm{Yb}_2\mathrm{Ti}_2\mathrm{O}_7$ can be written in a number of different basis, which is summarized in Table \ref{localglobalstandardexchangetable}. Many earlier works used the local basis, where the spin quantization axis is along the line connecting the center and corner of a tetrahedron unit. This was done based on the anticipation that the resulting spin model is an extended version of the local $\mathrm{XXZ}$ model, which promotes the quantum spin liquid with an emergent photon, often called the quantum spin ice.\cite{PhysRevB.69.064404,PhysRevLett.108.037202,PhysRevB.86.104412,1802.09198} In order for this to happen, the Ising part of the interaction must be dominant, which has been questioned in more recent experimental investigations.\cite{PhysRevLett.119.057203} Here we use a more conventional or standard representation, which allows us to write the spin model in terms of the familiar exchange interactions. Upon certain simplification, the spin model reduces to the nearest neighbor $JK\Gamma$ model on the pyrochlore lattice, where $J$ is the Heisenberg interaction, $K$ the Kitaev interaction, and $\Gamma$ the symmetric anisotropic exchange interaction. The main reason for this choice is that $K$ and $\Gamma$ are manifestly the dominant exchange interactions according to the experimentally determined parameters of the model (see Table \ref{localglobalstandardexchangetable}). Both $K$ and $\Gamma$ are highly anisotropic spin exchange interactions and are known to cause strong magnetic frustration. For example, the pure Kitaev model on the honeycomb lattice supports an exactly soluble quantum spin liquid ground state.\cite{KITAEV20062} Using the standard representation or basis, it becomes clear why the system is so frustrated or close to the classical phase boundary between two competing magnetically ordered phases.

In order to control the relative strength of quantum fluctuations and take into account both the semiclassical and extreme quantum limits, we introduce in our mean field theory a relative weight $r \in [0,1]$\cite{PhysRevB.69.035111} between the spin liquid and the magnetic order. Therefore, the total mean field Hamiltonian is given by $H^\mathrm{MF} = (1-r) H^\mathrm{MF}_\mathrm{SL} + r H^\mathrm{MF}_\mathrm{MO}$, where $H^\mathrm{MF}_\mathrm{SL}$ and $H^\mathrm{MF}_\mathrm{MO}$ are the mean field Hamiltonians of the quantum spin liquid and the classical magnetic order. We consider the $\mathbb{Z}_2$ uniform and the $U(1)$ monopole flux\cite{PhysRevB.79.144432} ansatzes as the possible quantum spin liquid ground states, as well as the splayed ferromagnet and the competing antiferromagnet for the classical magnetic orders. When $r=1$, we recover the classical limit, and when $r=0$, we obtain the quantum spin liquid ground state. Thus smaller $r$ means stronger quantum fluctuations.

The ambiguity in writing down the total mean field Hamiltonian allows possibly different values of $r$. In principle, $r$ should be determined dynamically, which is beyond the mean field description. In our work, we vary the value of $r$ and map out the phase diagram. When $r$ is finite, but close to $0$ ($1$), the pure quantum spin liquid (pure classical magnetic order) arises as the ground state. On the other hand, we find that there exists a window of intermediate values of $r$, where the coexisting phase or fractionalized magnetically ordered phase appears as a stable mean field ground state of the experimentally determined spin model. In this case, the spinon excitations represent strong quantum fluctuations and the overall magnitude of the magnetic order parameter is reduced. We then study the evolution of the phase diagram in the presence of an external magnetic field. We find that increased fields greatly suppress the quantum fluctuations or the spin liquid correlation. The coexisting phase disappears and only the conventional magnetically ordered states survive at sufficiently high fields.
%This is consistent with the experimental finding that the magnetically ordered ground state supports well-defined spin-wave excitations and look much more conventional in high magnetic field.

Our results demonstrate that the low lying excitation continuum observed in the recent inelastic neutron scattering experiment\cite{PhysRevLett.119.057203} on Yb$_2$Ti$_2$O$_7$ at weak magnetic fields may be attributed to deconfined spinons in the fractionalized magnetically ordered phase. The disappearance of the spin liquid/coexisting phase with sufficiently strong magnetic fields, which signals the complete confinement of spinons, is also consistent with the absence of such continuum and the presence of sharp magnon modes at high magnetic fields in the experiment. While we only tested two different spin liquid ansatzes, we have established the splayed ferromagnetic state with deconfined spinons as an alternative account of the experimental findings at the qualitative level.

The remainder of this paper is organized as follows. In Sec.~\ref{model}, we discuss the structure and symmetry of the pyrochlore lattice, and the spin model of $\mathrm{Yb}_2\mathrm{Ti}_2\mathrm{O}_7$. In Sec.~\ref{methodapproach}, we formulate the problem through the complex fermion mean field theory and the combination of spin liquid and magnetic Hamiltonians. The two spin liquid ansatzes under investigation are also introduced. In Sec.~\ref{result}, we show the phase diagram in the neighborhood of Gaulin and Coldea parametrizations, for different values of the weighting factor and the magnetic field. The spinon band structures and dynamical spin structure factors of the pure spin liquid and coexisting phases are then examined. In Sec.~\ref{discussion}, we summarize our work, and discuss possible improvements and implications to experiments.

\section{\label{model}Model}

\subsection{\label{structuresymmetry}Structure and Symmetry of Pyrochlore Lattice}

\begin{figure}
\includegraphics[scale=0.3]{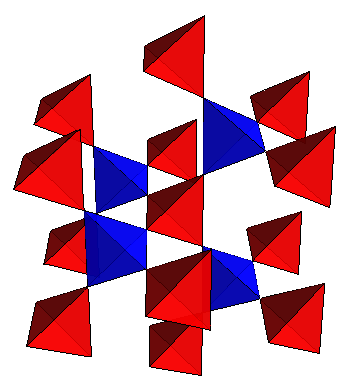}
\caption{\label{pyrochlorelatticefigure}The sites of pyrochlore lattice form a three dimensional network of corner sharing tetrahedra. The up (down) tetrahedra are colored in red (blue). It is easy to see that each up (down) tetrahedron is surrounded by four down (up) tetrahedra. The underlying Bravais lattice is the face centered cubic (fcc) lattice with four sites (sublattices) per unit cell, which are located at the corners of the tetrahedra.}
\end{figure}

\begin{figure}
\includegraphics[scale=0.3]{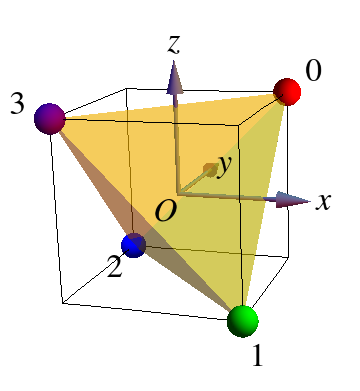}
\caption{\label{cubetetrahedronfigure}To visualize the tetrahedral space group $T_d$, we embed a tetrahedron in a cube and define a coordinate system with the cubic axes. The space group of the pyrochlore lattice is $\mathrm{Fd}\bar{3}\mathrm{m} = T_d \times \lbrace e, \mathcal{I} \rbrace$, where $e$ is the identity and $\mathcal{I}$ is inversion about a site.}
\end{figure}

Pyrochlore lattice is a three dimensional network of corner sharing tetrahedra (see Fig.~\ref{pyrochlorelatticefigure}). The underlying Bravais lattice is the face centered cubic (fcc) lattice, with four sites (or sublattices) per unit cell, which we label by $s=0$, $1$, $2$, and $3$. The space group of the pyrochlore lattice is $\mathrm{Fd} \bar{3} \mathrm{m}$,\cite{PhysRevB.79.144432} which is most conveniently viewed as $T_d \times i$,\cite{PhysRevB.95.094422} where $T_d$ is the tetrahedral symmetry group consisting of $24$ elements, and $i$ is the set containing identity $e$ and inversion $\mathcal{I}$ about a site. The elements of $T_d$ are best visualized by embedding the tetrahedron in a cube\cite{dresselhausgrouptheory,tinkhamgrouptheory} as in Fig.~\ref{cubetetrahedronfigure}: \\
\begin{center}
\begin{tabular}{rp{6.8 cm}}
$e$: & the identity; \\
$8$ $C_3$: & rotation by $\pm 2\pi/3$ about one of the local $[111]$ axes (the directions along the center to the corners of the tetrahedron); \\
$3$ $C_2$: & rotation by $\pi$ about one of the cubic axes ($x$, $y$ and $z$ directions); \\
$6$ $S_4$: & rotation by $\pm \pi/2$ about one of the cubic axes (e.g. $x$ axis) followed by reflection across the plane perpendicular to that axis (e.g. $yz$ plane); \\
$6$ $\sigma_\mathrm{d}$: & reflection across one of the diagonal planes, which are perpendicular to the $[011]$, $[01\bar{1}]$, $[101]$, $[\bar{1}01]$, $[110]$, and $[1\bar{1}0]$ directions.
\end{tabular}
\end{center}
In Fig.~\ref{cubetetrahedronfigure}, we have followed the choice of coordinates as in Ref.~\onlinecite{PhysRevX.1.021002}, such that the fcc Bravais lattice points are located at the centers of tetrahedra, and the sublattices $s=0$, $1$, $2$, and $3$ are displaced by $a/8 \left(1,1,1\right)$, $a/8 \left(1,-1,-1\right)$, $a/8 \left(-1,1,-1\right)$, and $a/8 \left(-1,-1,1\right)$ from the tetrahedral centers respectively, where $a$ is the lattice constant of the conventional cubic cell (which contains four fcc Bravais lattice points). The inversion center is chosen to be the sublattice $s=0$ in the unit cell at the origin $\mathbf{0}$.

\subsection{\label{spinhamiltoniansection}Spin Hamiltonian}
$\mathrm{Yb}_2 \mathrm{Ti}_2 \mathrm{O}_7$, a pyrochlore magnet with $J_\mathrm{eff}=1/2$ local moments (which are simply referred to as spins) residing on the corners of the tetrahedra, has long been considered as a candidate for quantum spin liquid. The most general nearest neighbor bilinear spin Hamiltonian 
\begin{equation} \label{spinhamiltonian}
H = \sum_{ij} \sum_{\mu \nu} S_i^\mu J_{ij}^{\mu \nu} S_j^\nu
\end{equation}
allowed by the symmetries of pyrochlore lattice contains four independent exchange parameters $J_1$, $J_2$, $J_3$, and $J_4$.\cite{PhysRevX.1.021002} These parameters are defined in the \textit{global} coordinates. $J_1$ is the Heisenberg interaction ($J$), $J_2-J_1$ the Kitaev interaction ($K$), $J_3$ the symmetric anisotropic exchange interaction ($\Gamma$), and $J_4$ the Dzyaloshinskii Moriya interaction ($D$). For instance, the interaction between the spins at sublattice $0$ and $1$ is given by,
\begin{equation} \label{spinhamiltonian01}
\begin{aligned}[b]
H_{01} &= \begin{pmatrix} S_0^x & S_0^y & S_0^z \end{pmatrix} \begin{pmatrix} J_2 & J_4 & J_4 \\ -J_4 & J_1 & J_3 \\ -J_4 & J_3 & J_1 \end{pmatrix} \begin{pmatrix} S_1^x \\ S_1^y \\ S_1^z \end{pmatrix} \\
&= J \mathbf{S}_0 \cdot \mathbf{S}_1 + K S_0^x S_1^x + \Gamma \left( S_0^y S_1^z + S_0^z S_1^y \right) \\ 
& \quad + D \left(S_0^x S_1^y - S_0^y S_1^x + S_0^x S_1^z - S_0^z S_1^x \right) .
\end{aligned}
\end{equation}
It is clear that $\langle 01 \rangle$ is an $x$ bond from the second equality. The interactions on other bonds can be obtained by symmetry,\cite{PhysRevX.1.021002,PhysRevB.95.094422} see Appendix \ref{localcoordinates}. It is also a common (arguably much more prevalent) practice to write the spin Hamiltonian \eqref{spinhamiltonian} in the \textit{local} coordinates,\cite{PhysRevX.1.021002} where the local $z$ axes are defined along the local $\left[ 111 \right]$ directions (see \eqref{localbases0}-\eqref{localbases3} in Appendix \ref{localcoordinates}),
\begin{equation} \label{spinhamiltonianlocal}
\begin{aligned}[b]
H &= \sum_{ij} \left[ J_{zz} \mathsf{S}_i^z \mathsf{S}_j^z - J_\pm \left( \mathsf{S}_i^+ \mathsf{S}_j^- + \mathsf{S}_i^- \mathsf{S}_j^+ \right) \right. \\
& \qquad \quad + J_{\pm \pm} \left( \gamma_{ij} \mathsf{S}_i^+ \mathsf{S}_j^+ + \gamma_{ij}^* \mathsf{S}_i^- \mathsf{S}_j^- \right) \\
& \qquad \quad \left. + J_{z \pm} \left( \zeta_{ij} \mathsf{S}_i^z \mathsf{S}_j^+ + \zeta_{ij}^* \mathsf{S}_i^z \mathsf{S}_j^- + i \longleftrightarrow j \right) \right].
\end{aligned}
\end{equation}
where $\gamma_{ij}$ and $\zeta_{ij}$ are unimodular complex numbers (see \eqref{zetamatrix} and \eqref{gammamatrix} Appendix \ref{localcoordinates}). In the form \eqref{spinhamiltonianlocal}, the spin Hamiltonian has the advantage that when the spin flip interactions are negligible, i.e.~in the limit $J_{\pm \pm} \longrightarrow 0$ and $J_{z \pm} \longrightarrow 0$, it reduces to a local $\mathrm{XXZ}$ model, which is studied in Refs.~\onlinecite{PhysRevB.69.064404,1802.09198} and shown to support quantum spin liquid states. The relation between local and global exchange parameters, $(J_{zz},J_{\pm},J_{\pm \pm},J_{z \pm})$ and $(J_1,J_2,J_3,J_4)$, can be found in \eqref{globallocalexchangerelation} in Appendix \ref{localcoordinates}.

The interaction parameters of the spin Hamilonian of $\mathrm{Yb}_2 \mathrm{Ti}_2 \mathrm{O}_7$ are obtained from spin wave analysis of inelastic neutron scattering at high magnetic fields.\cite{PhysRevX.1.021002,PhysRevLett.119.057203} We list the Gaulin and Coldea parametrizations of $\mathrm{Yb}_2 \mathrm{Ti}_2 \mathrm{O}_7$ in the local and global coordinates, as well as in the form of standard exchanges, in Table \ref{localglobalstandardexchangetable}. From the first row, we notice that $J_{\pm \pm}$ and/or $J_{z \pm}$ is comparable to, or much larger than, $J_{zz}$ and $J_{\pm}$. Moreover, $J_{zz}$ is not the largest energy scale, especially in the Coldea parametrization. There is thus no much merit to use the local coordinate system. In contrast, from the last row, we can easily see that $K$ and/or $\Gamma$ is the dominant interaction (with nonnegligible $J$ in Gaulin parametrization), which gives rise to strong frustration. Hence, it may be more convenient to work with the standard exchanges or in the global coordinates.

\begin{table}
\caption{\label{localglobalstandardexchangetable} Gaulin and Coldea parametrizations in the local $(J_{zz},J_{\pm},J_{\pm \pm},J_{z \pm})$ and global $(J_1,J_2,J_3,J_4)$ coordinates, and in the form of standard exchanges $(J,K,\Gamma,D)$. Energy is in units of $\mathrm{meV}$.}
\begin{ruledtabular}
\begin{tabular}{c|c|c}
 & Gaulin & Coldea \\ \hline
local & $(0.17,0.05,0.05,-0.14)$ & $(0.026,0.074,0.048,-0.159)$ \\
global & $(-0.09,-0.22,-0.29,0.01)$ & $(-0.028,-0.326,-0.272,0.049)$ \\
standard & $(-0.09,-0.13,-0.29,0.01)$ & $(-0.028,-0.298,-0.272,0.049)$ \\
\end{tabular}
\end{ruledtabular}
\end{table}

\begin{figure}
\includegraphics[scale=0.5]{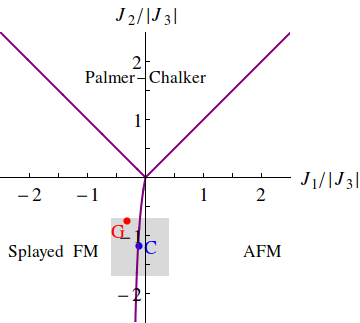}
\caption{\label{classicalphasediagramfigure}The classical phase diagram of the pyrochlore lattice in the $J_1 - J_2$ phase space (with $J_3=-1$) reported in Ref.~\onlinecite{PhysRevB.95.094422}. The location of Gaulin and Coldea parametrizations (with $J_4=0$), $(J_1,J_2)=(-0.31,-0.76)$ and $(-0.1,-1.2)$, are indicated. They are very close to the phase boundary between the splayed ferromagnetic (FM) and antiferromagnetic (AFM) orders. We investigate the phase diagram with the Hamiltonian \eqref{HMF}, which encodes both the spin liquid (quantum) and magnetically ordered (classical) phases, in the neighborhood of these parametrizations (the shaded area).}
\end{figure}

\begin{figure}
\subfloat[]{\label{splayedfmorderfigure} \includegraphics[scale=0.3]{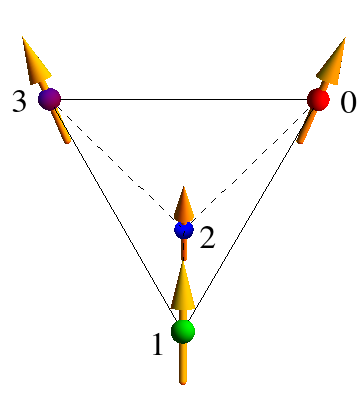}} \quad
\subfloat[]{\label{afmorderfigure} \includegraphics[scale=0.3]{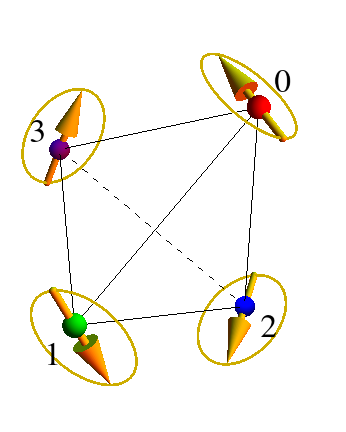}}
\caption{(a) The configuration of spins in the splayed ferromagnetic order. The spins on the four sublattices align in one of the cubic axes (e.g.~$z$ direction shown here) with some canting angles, which gives rise to a finite magnetization. (b) The configuration of spins in the antiferromagnetic order, which is a one dimensional manifold of states with zero net magnetization. The circular loops around the spins indicate the $U(1)$ symmetry. These figures are adapted from Ref.~\onlinecite{PhysRevB.95.094422}.}
\end{figure}

$J_4$ is negligible in Gaulin parametrization, though comparable to $J_1$ in Coldea parametrization. Still, it is one order of magnitude less than $J_2$ and $J_3$ in both cases. Therefore, to reduce the level of complexity we set $J_4=0$, so that the spin Hamiltonian \eqref{spinhamiltonian} is essentially the $J K \Gamma$ model,
\begin{subequations}
\begin{align}
H &= \sum_{\lambda=x,y,z} \sum_{\langle ij \rangle \in \lambda} \left( H_{ij}^J + H_{ij}^K + H_{ij}^\Gamma \right) ; \label{JKGmodel} \\
H_{ij}^J &= J \mathbf{S}_i \cdot \mathbf{S}_j , \label{heisenberg} \\
H_{ij}^K &= K S_i^\lambda S_j^\lambda , \label{kitaev} \\ 
H_{ij}^\Gamma &= \Gamma \left( S_i^\mu S_j^\nu + S_i^\nu S_j^\mu \right) \label{gamma} ,
\end{align}
\end{subequations}
where $(\lambda,\mu,\nu)$ is a cyclic permutation of $(x,y,z)$, on the pyrochlore lattice. Classically, both Gaulin and Coldea parametrizations lead to the splayed/noncollinear ferromagnetic (simply refered to as $\mathrm{FM}$) ground state, with a nearby competing antiferromagnetic ($\mathrm{AFM}$) phase (see Fig.~1 in Ref.~\onlinecite{PhysRevB.95.094422} and Fig.~S9 in the supplementary material of Ref.~\onlinecite{PhysRevLett.119.057203}). With $J_4$ set to $0$ and $J_3$ scaled to $-1$, we locate these parametrizations in the $J_1 - J_2$ phase space along with the magnetic orders derived in Ref.~\onlinecite{PhysRevB.95.094422}, as shown in Fig.~\ref{classicalphasediagramfigure}. We notice that Coldea parametrization falls into the $\mathrm{AFM}$ phase, but it is really an artifact of the simplification $J_4=0$. This happens because the full parametrization, while sitting on the $\mathrm{FM}$ side, is extremely close to the $\mathrm{FM}/\mathrm{AFM}$ boundary. Nevertheless, we will see later that a small magnetic field immediately stabilizes 
the FM phase for the simplified Coldea parametrization.

The FM phase has a finite magnetization along one of the cubic axes, from which the spins are canted away with certain angles that depend on the exchange couplings. The AFM phase has zero net magnetization, and possesses a $U(1)$ symmetry, i.e.~it is a one dimensional manifold of states with a continuous parameter. The spin configurations in these phases are depicted in Figs.~\ref{splayedfmorderfigure} and \ref{afmorderfigure}. It is shown in Ref.~\onlinecite{PhysRevB.95.094422} that the nearest neighbor bilinear spin model \eqref{spinhamiltonian} on the pyrochlore lattice admits only $\mathbf{q}=\mathbf{0}$ orderings, i.e.~all the possible symmetry breaking patterns are invariant under a Bravais lattice translation. Therefore, it is sufficient to know the arrangement of spins on the four sublattices of a tetrahedron.

In the presence of an external magnetic field $\mathbf{B}$, the term $- \mu_\mathrm{B} B^ \mu g^{\mu \nu} S^\nu$ is added to the spin Hamiltonian \eqref{spinhamiltonian}. In the \textit{local} coordinates, the $g$ tensor takes the form\cite{PhysRevX.1.021002,PhysRevB.95.094422}
\begin{equation}
g^{\mu \nu}_\mathrm{local} = \begin{pmatrix} g_{xy} & 0 & 0 \\ 0 & g_{xy} & 0 \\ 0 & 0 & g_z\end{pmatrix} . \label{localgtensor}
\end{equation}
The $g$ tensor in the global coordinates can be obtained by suitable rotations of \eqref{localgtensor}, whose expression can be found in \eqref{gtensorglobal} in Appendix \ref{localcoordinates}. From now on we will absorb the Bohr magneton factor into the magnetic field, $\mu_\mathrm{B} \mathbf{B} \longrightarrow \mathbf{B}$, so that it has the same unit as energy.

\section{\label{methodapproach}Method and Approach}

\subsection{\label{complexfermionmeanfieldtheory}Complex Fermion Mean Field Theory}
We first represent the spin operator in terms of fermionic spinon 
creation and annihilation operators,
\begin{equation} \label{secondquantizationspinoperator}
\mathbf{S}_i = \frac{1}{2} f_{i \alpha}^\dagger \mathbf{\sigma}_{\alpha \beta} f_{i \beta}.
\end{equation}
The spin Hamiltonian \eqref{spinhamiltonian} is then quartic in these spinon operators. We also define the bond operators\cite{PhysRevB.80.064410}
\begin{subequations}
\begin{align}
\hat{\chi}_{ij} &= \sum_{\alpha} f_{i \alpha}^\dagger f_{j \alpha} , \label{singlethopping} \\
\hat{\Delta}_{ij} &= \sum_{\alpha \beta} f_{i \alpha} [i \sigma^y]_{\alpha \beta} f_{j \beta} , \label{singletpairing} \\
\hat{E}^\mu_{ij} &= \sum_{\alpha \beta} f_{i \alpha}^\dagger \sigma^\mu_{\alpha \beta} f_{j \beta} , \label{triplethopping} \\
\hat{D}^\mu_{ij} &= \sum_{\alpha \beta} f_{i \alpha} [i \sigma^y \sigma^\mu]_{\alpha \beta} f_{j \beta} , \label{tripletpairing}
\end{align}
\end{subequations}
where $\mu=x,y,z$, which describe the singlet hopping, singlet pairing, triplet hopping, and triplet pairing of spinons at site $i$ and $j$ respectively. If we are only concerned with the spin liquid state (deconfined spinons), we can express the Hamiltonian of a generic nearest neighbor $JK\Gamma$ model \eqref{JKGmodel} in terms of the bond operators \eqref{singlethopping}-\eqref{tripletpairing} as
\begin{widetext}
\begin{subequations}
\begin{align}
H_\mathrm{SL} &= \sum_{\lambda=x,y,z} \sum_{\langle ij \rangle \in \lambda} \left( H_{ij}^J + H_{ij}^K + H_{ij}^\Gamma \right) + \mu_3 \sum_{i \alpha} \left( f_{i \alpha}^\dagger f_{i \alpha} - 1 \right) + \left( (\mu_1 + i \mu_2) \sum_i  f_{i \downarrow} f_{i \uparrow}  + \mathrm{h.c.} \right) ; \label{HSL} \\
H_{ij}^J &= \left \lbrace \begin{array}{l} - \dfrac{\lvert J \rvert}{4} \left( \hat{\mathbf{E}}_{ij}^\dagger \cdot \hat{\mathbf{E}}_{ij} + \hat{\mathbf{D}}_{ij}^\dagger \cdot \hat{\mathbf{D}}_{ij} \right) , \, \mathrm{for} \, J<0; \\
- \dfrac{\lvert J \rvert}{4} \left( \hat{\chi}_{ij}^\dagger \hat{\chi}_{ij} + \hat{\Delta}_{ij}^\dagger \hat{\Delta}_{ij} \right) , \, \mathrm{for} \, J>0; \end{array} \right. \label{heisenbergbondoperator} \\
H_{ij}^K &= \left \lbrace \begin{array}{l} - \dfrac{\lvert K \rvert}{8} \left( \hat{E}^{\mu \dagger}_{ij} \hat{E}^{\mu}_{ij} + \hat{E}^{\nu \dagger}_{ij} \hat{E}^{\nu}_{ij} + \hat{D}^{\mu \dagger}_{ij} \hat{D}^{\mu}_{ij} + \hat{D}^{\nu \dagger}_{ij} \hat{D}^{\nu}_{ij} \right) , \, \mathrm{for} \, K<0; \\
- \dfrac{\lvert K \rvert}{8} \left( \hat{\chi}_{ij}^\dagger \hat{\chi}_{ij} + \hat{\Delta}_{ij}^\dagger \hat{\Delta}_{ij} + \hat{E}^{\lambda \dagger}_{ij} \hat{E}^{\lambda}_{ij} + \hat{D}^{\lambda \dagger}_{ij} \hat{D}^{\lambda}_{ij} \right) , \, \mathrm{for} \, K>0; \end{array} \right. \label{kitaevbondoperator} \\
H_{ij}^\Gamma &= \left \lbrace \begin{array}{l} - \dfrac{\lvert \Gamma \rvert}{8} \left( \left( \hat{E}^{\mu}_{ij} - \hat{E}^{\nu}_{ij} \right)^\dagger \left( \hat{E}^{\mu}_{ij} - \hat{E}^{\nu}_{ij} \right) + \left( \hat{D}^{\mu}_{ij} - \hat{D}^{\nu}_{ij} \right)^\dagger \left( \hat{D}^{\mu}_{ij} - \hat{D}^{\nu}_{ij} \right) + \hat{\chi}_{ij}^\dagger \hat{\chi}_{ij} + \hat{\Delta}_{ij}^\dagger \hat{\Delta}_{ij} + \hat{E}^{\lambda \dagger}_{ij} \hat{E}^{\lambda}_{ij} + \hat{D}^{\lambda \dagger}_{ij} \hat{D}^{\lambda}_{ij} \right) , \, \mathrm{for} \, \Gamma<0; \\
- \dfrac{\lvert \Gamma \rvert}{8} \left( \left( \hat{E}^{\mu}_{ij} + \hat{E}^{\nu}_{ij} \right)^\dagger \left( \hat{E}^{\mu}_{ij} + \hat{E}^{\nu}_{ij} \right) + \left( \hat{D}^{\mu}_{ij} + \hat{D}^{\nu}_{ij} \right)^\dagger \left( \hat{D}^{\mu}_{ij} + \hat{D}^{\nu}_{ij} \right) + \hat{\chi}_{ij}^\dagger \hat{\chi}_{ij} + \hat{\Delta}_{ij}^\dagger \hat{\Delta}_{ij} + \hat{E}^{\lambda \dagger}_{ij} \hat{E}^{\lambda}_{ij} + \hat{D}^{\lambda \dagger}_{ij} \hat{D}^{\lambda}_{ij} \right) , \, \mathrm{for} \, \Gamma>0. \end{array} \right. \label{gammabondoperator}
\end{align}
\end{subequations}
\end{widetext}
The Lagrange multipliers $\mu_1,\mu_2,\mu_3 \in \mathbb{R}$ are introduced in \eqref{HSL} to enforce the single occupancy constraint (one spinon per site) on average. Note that we have carefully written the various interactions \eqref{heisenbergbondoperator}-\eqref{gammabondoperator} in the form
\begin{equation}
H_{ij}^X = - \lvert c^X \rvert \sum_O \hat{O}_{ij}^\dagger \hat{O}_{ij}, \label{quadraticform}
\end{equation}
from which a mean field decoupling naturally follows,
\begin{equation}
H_{ij}^X = - \lvert c^X \rvert \sum_O \left( O_{ij}^* \hat{O}_{ij} + O_{ij} \hat{O}_{ij}^\dagger - \lvert O_{ij} \rvert^2 \right) , \label{meanfielddecoupling}
\end{equation}
and the mean field energy is bounded below (i.e.~the stability requirement is satisfied). $\hat{O}_{ij}$ are the bond operators \eqref{singlethopping}-\eqref{tripletpairing} as before, while $O_{ij}$ (without the hat) are variational parameters to minimize the mean field energy. We denote the spin liquid Hamiltonian \eqref{HSL} after the mean field decoupling \eqref{meanfielddecoupling} as $H_\mathrm{SL}^\mathrm{MF}$.

However, $H_\mathrm{SL}^\mathrm{MF}$ tells us nothing about the classically ordered phases. To capture these phases, we make use of the result of Ref.~\onlinecite{PhysRevB.95.094422}, which provides a group theory analysis of the classical model (i.e.~the spins in the Hamiltonian \eqref{spinhamiltonian} are treated as three component vectors with fixed magnitude), and lists all the possible magnetic orders with the corresponding order parameters. It is shown that the spin interactions \eqref{spinhamiltonian} on a tetrahedron can be expressed as a summation of bilinears of the order parameters $\mathbf{m}_\mathrm{X}$,\cite{PhysRevB.95.094422,1742-6596-145-1-012032} each of which is a linear combination of the components of the spins residing at the four sublattices, multiplied by some energy coefficients $a_\mathrm{X}$, each of which is a linear combination of the exchange couplings, as follows,
\begin{equation}
H_\mathrm{MO}^\mathrm{tet} = \frac{1}{2} \sum_\mathrm{X} a_\mathrm{X} \lvert \mathbf{m}_\mathrm{X} \rvert^2 \label{HMOtet}
\end{equation}
Since each unit cell contains one up and one down tetrahedra, the total magnetic Hamiltonian $H_\mathrm{MO}$ is given by summing $2$ times \eqref{HMOtet} over the unit cells. Furthermore, we keep only the $\mathrm{FM}$ ($\mathrm{X}=\mathrm{T}_{1,\mathrm{A}'}$) and $\mathrm{AFM}$ ($\mathrm{X}=\mathrm{E}$) order parameters as they are the only relevant classical phases to $\mathrm{Yb}_2 \mathrm{Ti}_2 \mathrm{O}_7$ while setting others to zero. Interested readers can refer to Table $\mathrm{III}$ and $\mathrm{V}$ in Ref.~\onlinecite{PhysRevB.95.094422} for the expressions of the order parameters and the energy coefficients of the various magnetic phases, here we only quote those of the $\mathrm{FM}$ and $\mathrm{AFM}$ phases,
\begin{widetext}
\begin{subequations}
\begin{align}
a_{\mathrm{T}_{1,\mathrm{A}'}} &= \left( 2 J_1 + J_2 \right) \cos^2 \theta_{\mathrm{T}_1} - (J_2 + J_3 - 2 J_4) \sin^2 \theta_{\mathrm{T}_1} + \sqrt{2} J_3 \sin 2 \theta_{\mathrm{T}_1} \label{FMcoefficient} , \\
\mathbf{m}_{\mathrm{T}_{1,\mathrm{A}'}} &= \frac{1}{2} \cos \theta_{\mathrm{T}_1} \begin{pmatrix} S_0^x + S_1^x + S_2^x + S_3^x \\ S_0^y + S_1^y + S_2^y + S_3^y \\ S_0^z + S_1^z + S_2^z + S_3^z \end{pmatrix} + \frac{1}{2\sqrt{2}} \sin \theta_{\mathrm{T}_1} \begin{pmatrix} S_0^y + S_0^z - S_1^y - S_1^z - S_2^y + S_2^z + S_3^y - S_3^z \\ S_0^z + S_0^x + S_1^z - S_1^x - S_2^z - S_2^x - S_3^z + S_3^x \\ S_0^x + S_0^y - S_1^x + S_1^y + S_2^x - S_2^y - S_3^x - S_3^y \end{pmatrix} , \label{FMOP} \\
\theta_{\mathrm{T}_1} &= \frac{1}{2} \tan^{-1} \left( \frac{\sqrt{8} J_3}{2 J_1 + 2 J_2 + J_3 - 2 J_4} \right) ; \label{thetaT1definition} \\
a_\mathrm{E} &= - 2 J_1 + J_2 + J_3 + 2 J_4 , \label{AFMcoefficient} \\
\mathbf{m}_\mathrm{E} &= \begin{pmatrix} \dfrac{1}{2 \sqrt{6}} \left( - 2 S_0^x + S_0^y + S_0^z - 2 S_1^x - S_1^y - S_1^z + 2 S_2^x + S_2^y - S_2^z + 2 S_3^x - S_3^y + S_3^z \right) \\ \dfrac{1}{2 \sqrt{2}} \left( - S_0^y + S_0^z + S_1^y -S_1^z - S_2^y - S_2^z + S_3^y + S_3^z \right) \end{pmatrix} . \label{AFMOP}
\end{align}
\end{subequations}
\end{widetext}
We represent the magnetic order parameters in terms of spinon operators using \eqref{secondquantizationspinoperator}, and carry out a mean field decoupling similar to \eqref{meanfielddecoupling},
\begin{equation}
H_\mathrm{MO}^\mathrm{MF} = \sum_\mathbf{R} \sum_X a_\mathrm{X} \left( 2 \mathbf{m}_\mathrm{X} \cdot \hat{\mathbf{m}}_\mathrm{X} - \lvert \mathbf{m}_\mathrm{X} \rvert^2 \right) , \label{HMOMF}
\end{equation}
where $\mathbf{R}$ labels the unit cell (not individual site), and $\mathbf{m}_\mathrm{X}$ (without the hat) are now variational parameters. The stability requirement is satisfied as the coefficients $a_{\mathrm{T}_{1,\mathrm{A}'}}$ and $a_\mathrm{E}$ are negative in the $J_1-J_2$ phase space (with $J_3=-1$ fixed) under study (see Fig.~\ref{classicalphasediagramfigure}). This allows us to incorporate both the quantum spin liquid and magnetically ordered states into a single Hamiltonian
\begin{equation}
H^\mathrm{MF} = (1-r) H_\mathrm{SL}^\mathrm{MF} + r H_\mathrm{MO}^\mathrm{MF} . \label{HMF}
\end{equation}
However, there is an ambiguity in combining the two Hamiltonians $H_\mathrm{SL}^\mathrm{MF}$ and $H_\mathrm{MO}^\mathrm{MF}$, which is reflected in the introduction of the weighting factor $r$ in \eqref{HMF}. In principle, $r$ can assume any values from $0$ (pure spin liquid description, quantum limit) to $1$ (pure magnetic order description, classical limit). If one takes $r=1/2$, which seems to be an intuitive choice, a self consistent calculation always drives the system to the purely classical magnetically ordered phase with all the spin liquid parameters (i.e.~the spinon hoppings and pairings) converging to zero. 
In order to incorporate quantum fluctuations, somehow we have to suppress the classical order by further decreasing the value of $r$ from $1/2$.
Such a scheme, for example, was applied in the previous mean field study of the Kondo-Heisenberg model.\cite{PhysRevB.69.035111}
In principle, the value of $r$ would be determined dynamically if the fluctuations beyond mean field theory could be incorporated. 
At present, there is no systematic way to determine which value of $r$ should be used within a mean field theory. 
In this work, we will vary the value of $r$ and investigate how the phase diagram evolves with respect to decreasing $r$. 
In particular, we investigate whether there exist reasonable values of $r$ for which the spin liquid coexists with a magnetic order (i.e.~magnetic order with fractionalized excitations) in the neighborhood of Gaulin and Coldea parametrizations of $\mathrm{Yb}_2 \mathrm{Ti}_2 \mathrm{O}_7$. Certainly a vanishingly small value of $r$ is not good, as in this case the magnetic order is completely suppressed and the spin liquid phase is always obtained. We will find that, for the spin liquid ansatzes in Section \ref{spinliquidansatzes}, when $r$ is decreased to $\sim 0.25$, a coexisting phase of spin liquid and magnetic order can be stabilized. 

The self consistent equations for all the variational parameters are obtained by minimizing the mean field Hamiltonian \eqref{HMF},
\begin{equation}
\begin{aligned}[b]
& \frac{\partial \langle H^\mathrm{MF} \rangle}{\partial O} = 0 \Longleftrightarrow O = \langle \hat{O} \rangle, \\
& O = \chi_{ij}, \Delta_{ij}, \mathbf{E}_{ij}, \mathbf{D}_{ij}, \mathbf{m}_{\mathrm{T}_{1,\mathrm{A}'}}, \mathbf{m}_\mathrm{E}, \label{selfconsistentequations}
\end{aligned}
\end{equation}
while the Lagrange multipliers $\mu_1, \mu_2, \mu_3$ are chosen such that the single occupancy constraint
\begin{equation}
\sum_{\alpha} f_{i \alpha}^\dagger f_{i \alpha} = 1 \label{singleoccupancy}
\end{equation}
is satisfied on average. The self consistent calculations are performed in momentum space, through the Fourier transfrom
\begin{equation}
f_{\mathbf{k},s,\alpha} = \frac{1}{\sqrt{N}} \sum_\mathbf{R} f_{\mathbf{R},s,\alpha} e^{-i \mathbf{k} \cdot \mathbf{R}} , \label{fouriertransform}
\end{equation}
where $\mathbf{R}$, $s$ and $\alpha$ label the unit cell, sublattice, and spin flavor respectively.

\subsection{\label{spinliquidansatzes}Spin Liquid Ansatzes}

\begin{figure*}
\subfloat[$r=0.25,B_z/\lvert J_3 \rvert=0$.]{\label{Z2Ur025b000figure} \includegraphics[scale=0.3]{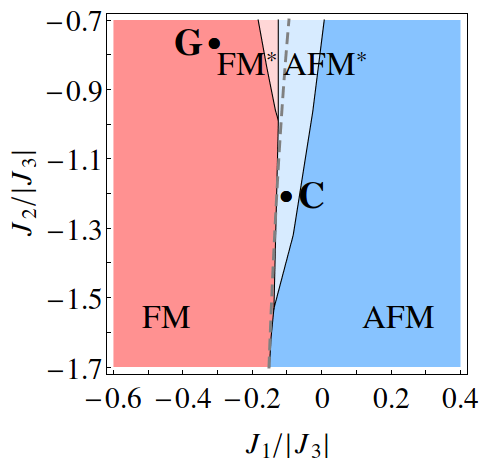}} \quad
\subfloat[$r=0.23,B_z/\lvert J_3 \rvert=0$.]{\label{Z2Ur023b000figure} \includegraphics[scale=0.3]{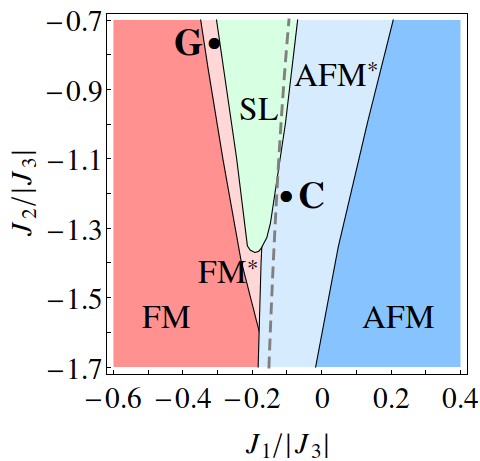}} \quad
\subfloat[$r=0.20,B_z/\lvert J_3 \rvert=0$.]{\label{Z2Ur020b000figure} \includegraphics[scale=0.3]{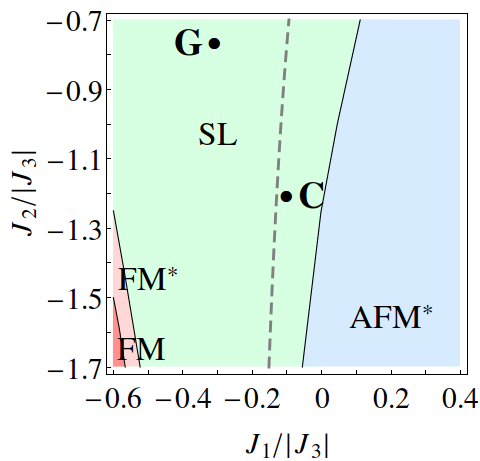}} \\
\subfloat[$r=0.23,B_z/\lvert J_3 \rvert=0.01$.]{\label{Z2Ur023b001figure} \includegraphics[scale=0.3]{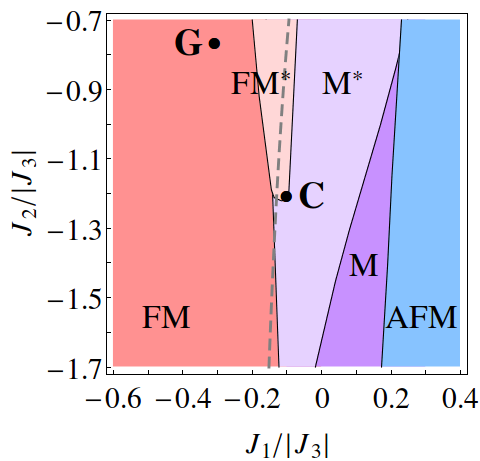}} \quad
\subfloat[$r=0.23,B_z/\lvert J_3 \rvert=0.02$.]{\label{Z2Ur023b002figure} \includegraphics[scale=0.3]{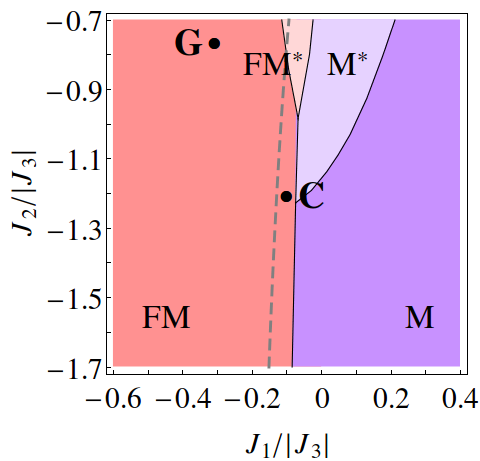}} \quad
\subfloat[$r=0.23,B_z/\lvert J_3 \rvert=0.04$.]{\label{Z2Ur023b004figure} \includegraphics[scale=0.3]{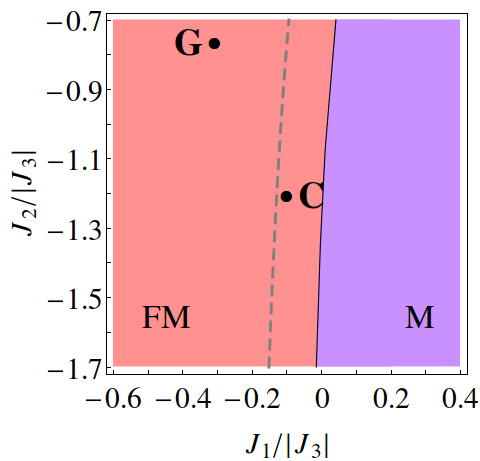}}
\caption{The $J_1 - J_2$ phase diagrams (with $J_3=-1$) of the model \eqref{HMF} with the $\mathbb{Z}_2 \mathrm{U}$ ansatz. The gray dashed line represents the classical phase boundary (see Fig.~\ref{classicalphasediagramfigure}). The phase diagrams at zero magnetic field, with weighting factors (a) $r=0.25$, (b) $r=0.23$, and (c) $r=0.20$. The area with deconfined spinons (either a pure spin liquid or coexisting phase) expands as $r$ decreases due to the suppression of classical order. Choosing a representative weighting factor $r=0.23$, we apply magnetic fields (d) $B_z=0.01 \lvert J_3 \rvert$, (e) $B_z=0.02 \lvert J_3 \rvert$, and (f) $B_z=0.04 \lvert J_3 \rvert$ along one of the cubic axes. The area with deconfined spinons shrinks and eventually disappears with increasing field. The meaning of the various labels can be found in the main text, particularly in Sec. \ref{phasediagram}.}
\end{figure*}

\begin{figure*}
\subfloat[$r=0.25,B_z/\lvert J_3 \rvert=0$.]{\label{U1Mr025b000figure} \includegraphics[scale=0.3]{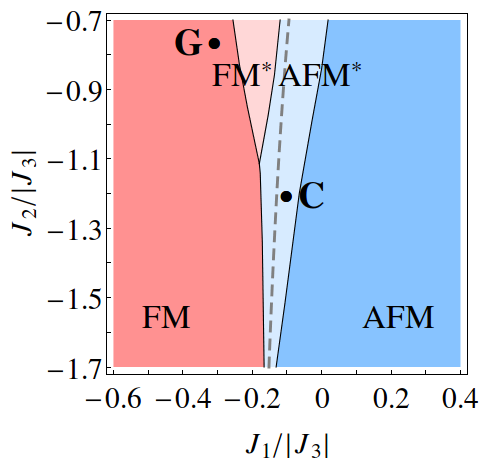}} \quad
\subfloat[$r=0.23,B_z/\lvert J_3 \rvert=0$.]{\label{U1Mr023b000figure} \includegraphics[scale=0.3]{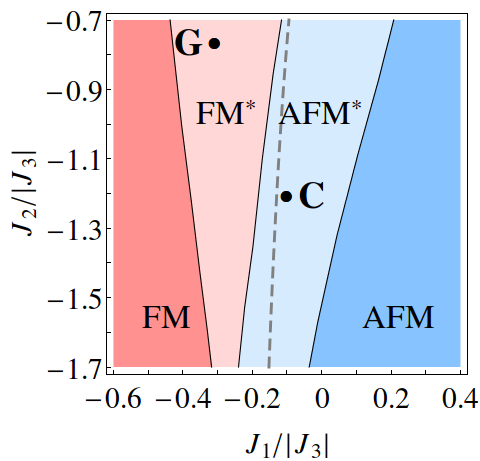}} \quad
\subfloat[$r=0.20,B_z/\lvert J_3 \rvert=0$.]{\label{U1Mr020b000figure} \includegraphics[scale=0.3]{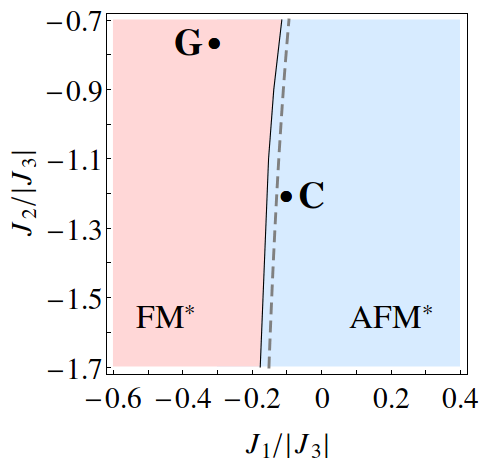}} \\
\subfloat[$r=0.23,B_z/\lvert J_3 \rvert=0.01$.]{\label{U1Mr023b001figure} \includegraphics[scale=0.3]{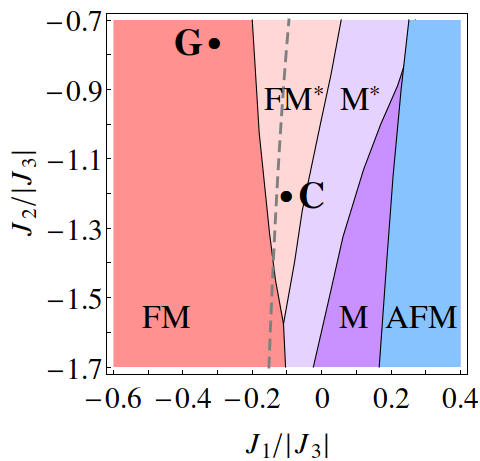}} \quad
\subfloat[$r=0.23,B_z/\lvert J_3 \rvert=0.02$.]{\label{U1Mr023b002figure} \includegraphics[scale=0.3]{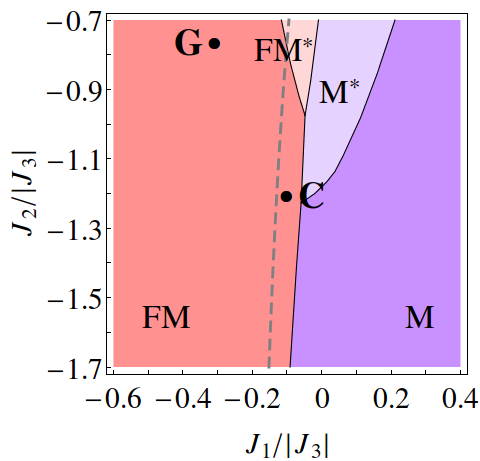}} \quad
\subfloat[$r=0.23,B_z/\lvert J_3 \rvert=0.04$.]{\label{U1Mr023b004figure} \includegraphics[scale=0.3]{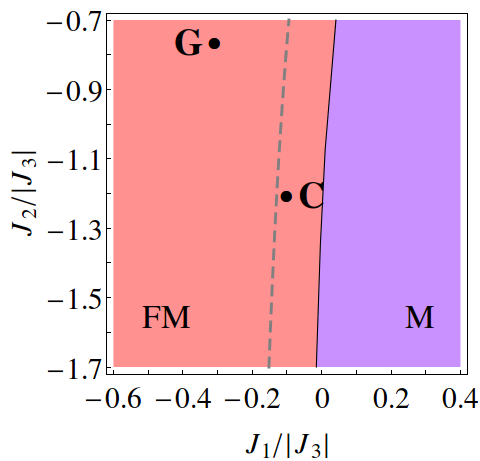}}
\caption{The $J_1 - J_2$ phase diagrams (setting $J_3=-1$) of the model \eqref{HMF} with the $U(1) \mathrm{M}$ ansatz, at various weighting factors $r$ and magnetic fields $B_z$ studied above. The main difference between the $U(1) \mathrm{M}$ and $\mathbb{Z}_2 \mathrm{U}$ ansatzes is that no pure spin liquid state appears in the phase diagram, as the magnetic order parameter always converges to some finite number, although it can be as small as $\lesssim 0.01$ of its classical value (compare (a)-(c) here to Figs.~\ref{Z2Ur025b000figure}-\ref{Z2Ur020b000figure}). The qualitative features which remain the same are that the area with deconfined spinons expands as $r$ decreases due to the suppression of classical order, and shrinks with increasing magnetic field $B_z$.}
\end{figure*}

A specific choice of the set of spinon hopping and pairing parameters $\lbrace \chi_{ij} , \Delta_{ij}, \mathbf{E}_{ij},\mathbf{D}_{ij} \rbrace$ is called a mean field ansatz of the spin liquid. The representation of spin operator by spinons \eqref{secondquantizationspinoperator} introduces an $SU(2)$ gauge redundancy
\begin{equation}
\Psi_i \longrightarrow \Psi_i G_i, \, \Psi_i = \begin{pmatrix} f_{i \uparrow} & f_{i \downarrow} \\ f_{i \downarrow}^\dagger & -f_{i \uparrow}^\dagger \end{pmatrix}, \, G_i \in SU(2) . \label{gaugeredundancy}
\end{equation}
Consequently, the various symmetries of the system (space group and time reversal) can be realized \textit{projectively} at the mean field level. That is, for the Hamiltonian $H_\mathrm{SL}^\mathrm{MF}$ to be invariant under a symmetry transformation $X$, the mean field ansatz should respect $X$ up to a gauge transformation $G_X$. The collection of the compound operators $G_X X$, which leaves the mean field ansatz unchanged, is known as the projective symmetry group (PSG).\cite{PhysRevB.65.165113} PSG classification enables one to enumerate all the different mean field ansatzes (distinguished by {$G_X$}) consistent with the symmetries of the system. Readers are encouraged to refer to Refs.~\onlinecite{PhysRevB.65.165113,PhysRevB.83.224413,PhysRevB.95.054404} for more details on PSG.

Nevertheless, we do not consider a complete PSG classification in this paper due to the reasons below. First, the pyrochlore lattice is a highly symmetric three dimensional structure, so that PSG classification is likely to result in a large number of mean field ansatzes. It is then impractical to examine their physical properties (energy, band structure, phase diagram, etc.) exhaustively. Second, our focus is to demonstrate that it is possible to open up a spin liquid/coexisting phase by taking into account some amount of quantum fluctuations (i.e.~choosing a weighting factor r that is not too small) in the $J_1-J_2$ phase space near the experimentally determined parametrizations of $\mathrm{Yb}_2 \mathrm{Ti}_2 \mathrm{O}_7$. For this purpose, we will only study two simple ansatzes, the $\mathbb{Z}_2$ uniform spin liquid ansatz and the $U(1)$ monopole flux spin liquid ansatz, which are simply refered to as $\mathbb{Z}_2 \mathrm{U}$ and $U(1) \mathrm{M}$ respectively.

In the $\mathbb{Z}_2$ uniform spin liquid state, the space group of the pyrochlore lattice and the time reversal symmetry are both preserved, and these symmetries are realized trivially (i.e.~for any symmetry element $X$, the associated gauge transformation $G_X=1$ is trivial). This state has four independent spin liquid parameters $\chi_{01}$, $\Delta_{01}$, $E^y_{01}$, and $D^y_{01}$, which are the spinon hoppings and pairings on the bond $\langle 01 \rangle$, and to which those at other bonds can be related by symmetry. For more details, see Appendix \ref{Z2Uansatzdetails}.

On the other hand, the monopole flux state is characterized by the $\pi/2$ static gauge flux (which is defined as the sum of the phases of the singlet hopping amplitudes $\chi_{ij}$ around a closed loop) that pierces each triangular face of a tetrahedron. This flux configuration can be obtained by placing a monopole of strength $2\pi$ inside each tetrahedron. This state is first constructed in Ref.~\onlinecite{PhysRevB.79.144432}. Not all symmetry elements are preserved in the monopole flux state. Time reversal symmetry is broken because the flux through a triangle is not $0$ or $\pi$. Only half of the $48$ pyrochlore space group elements are realized (projectively), while the other half consisting of inversion, reflections, and improper rotations are broken.\cite{PhysRevB.79.144432} A simple PSG is devised where the gauge transformation $G_X$ associated with the symmetry element $X$ is just $\pm 1$. Furthermore, the monopole flux state is a $U(1)$ spin liquid because no pairing terms are considered in the mean field Hamiltonian. It is found to be the lowest energy state of the nearest neighbor antiferromagnetic Heisenberg model on the pyrochlore lattice among the six candidates considered in Ref.~\onlinecite{PhysRevB.79.144432}. Singlet hopping $\chi_{ij}$ is the sole mean field parameter in that case. For the generic $JK\Gamma$ model, which contains anisotropic spin interactions, both singlet and triplet channels are present in the Hamiltonian (see \eqref{HSL}-\eqref{gammabondoperator}). Extending the monopole flux state to our Hamiltonian $H_\mathrm{SL}^\mathrm{MF}$, we keep the form of $\chi_{ij}$ as in Ref.~\onlinecite{PhysRevB.79.144432}, while constraining that of $\mathbf{E}_{ij}$ with the monopole flux PSG. All the spinon pairings are set to zero. There are four independent spin liquid parameters $\chi_{01}$, $E^x_{01}$, $E^y_{01}$, and $E^z_{01}$. For more details, see Appendix \ref{U1Mansatzdetails}.

\section{\label{result}Result}

\subsection{\label{phasediagram}Phase Diagram}

For the $\mathbb{Z}_2\mathrm{U}$ ansatz, when the weighting factor $r$ multiplying the magnetic Hamiltonian $H^\mathrm{MF}_\mathrm{MO}$ in \eqref{HMF} is decreased to about $0.25$, we can stabilize a coexisting phase, where both the spin liquid and magnetic order parameters are finite upon convergence of the self consistent equations, over a finite area in the $J_1-J_2$ phase space (see Fig.~\ref{Z2Ur025b000figure}). We label such a phase by $\mathrm{FM}^*$ or $\mathrm{AFM}^*$ depending on which magnetic order parameter is turned on, which means `a magnetically ordered state with fractionalized excitations/deconfined spinons'. As $r$ is further decreased, for instance to $0.23$ and $0.20$, magnetic ordering is further suppressed, the phase region with deconfined spinons expands, and a pure spin liquid phase, where all the magnetic order parameters converge to zero, emerges (see Figs.~\ref{Z2Ur023b000figure} and \ref{Z2Ur020b000figure}).

In the classical model, with the normalization $\lvert \mathbf{S}_i \rvert = 1/2$, we have the identity $\sum_\mathrm{X} \lvert \mathbf{m}_\mathrm{X} \rvert^2 =1$, and the maximum norm that each of the magnetic order parameters $\mathbf{m}_\mathrm{X}$ can reach is $1$. It is shown in Ref.~\onlinecite{PhysRevB.95.094422} that, in the absence of external magnetic field, $\lvert \mathbf{m}_\mathrm{X} \rvert = 1$ in the classically ordered phase $\mathrm{X}$ and $\lvert \mathbf{m}_\mathrm{Y} \lvert = 0$ for all other $\mathrm{Y} \neq \mathrm{X}$, in order to minimize the total energy. This is also true in our model, when all spin liquid parameters converge to zero, then $\lvert \mathbf{m}_\mathrm{T_{1,\mathrm{A}'}} \rvert = 1$ ($\lvert \mathbf{m}_\mathrm{E} \rvert = 1$) and $\lvert \mathbf{m}_\mathrm{E} \rvert = 0$ ($\lvert \mathbf{m}_\mathrm{T_{1,\mathrm{A}'}} \rvert = 0$) in the $\mathrm{FM}$ ($\mathrm{AFM}$) phase. However, when some spin liquid parameters are finite, then the norm of the magnetic order parameter does not attain its saturated value, i.e.~$\lvert \mathbf{m}_\mathrm{X} \rvert<1$ while all other $\lvert \mathbf{m}_\mathrm{Y \neq X} \rvert=0$. Since the magnetic order parameter is a linear combination of spin components, the magnitude of the expectation value of the spin operator $S \equiv \lvert \langle \hat{\mathbf{S}} \rangle \rvert$ decreases accordingly from the normalization $S_0=1/2$ in the presence of deconfined spinons. We can thus use the ratio $S/S_0$ to represent the reduction of the magnetic order parameter relative to its maximum norm. The advantage of considering $S/S_0$ instead of individual $\mathbf{m}_\mathrm{X}$ is that, as we shall see later, multiple magnetic order parameters can be simultaneously finite upon turning on an external magnetic field, while $\sum_\mathrm{X} \lvert \mathbf{m}_\mathrm{X} \rvert^2 < 1$ due to quantum fluctuations.

For the $U(1)\mathrm{M}$ ansatz, we can similarly obtain the coexisting phases $\mathrm{FM}^*$ and $\mathrm{AFM}^*$ at $r \sim 0.25$ (see Fig.~\ref{U1Mr025b000figure}). However, in these phases, $S/S_0 \sim 0.01$, leading to a small but finite magnetic order parameter. In contrast, for the $\mathbb{Z}_2U$ ansatz, $S/S_0$ is usually of the order of $0.1$ in the coexisting phase. Interestingly, decreasing $r$ further to $0.23$ and $0.20$, the area in the phase space with deconfined spinons expands (see Figs.~\ref{U1Mr023b000figure} and \ref{U1Mr020b000figure}), but always with a finite magnetic order parameter, whose magnitude can be as small as $\lesssim 0.01$ of the classical value (see Table \ref{U1Mgaulintable} in Appendix \ref{localglobalcomparison}, for example). Strictly speaking, no pure spin liquid phase is obtained in this case, but one can say that the coexisting phase obtained with the $U(1)\mathrm{M}$ ansatz is 
almost a pure spin liquid due to extremely small magnetic order parameter.

We pick a representative value of the weighting factor $r=0.23$ and investigate the evolution of phase diagram with the application of magnetic field along one of the cubic axes (in the $z$ direction, say). We fix the $g$ factors $g_{xy}=4.2$ and $g_z=2.0$ in \eqref{localgtensor}, based on the reported values of $(g_{xy},g_z)=(4.27,1.79)$ and $(4.17,2.14)$ in Refs.~\onlinecite{0953-8984-13-41-318,PhysRevLett.119.057203} respectively. The $\mathrm{FM}$ phase is energetically favored under such a field. With increasing field strength, we observe that the phase region with deconfined spinons shrinks, while that of $\mathrm{FM}$ grows and crosses the classical phase boundary at zero field (see Figs.~\ref{Z2Ur023b001figure} and \ref{Z2Ur023b002figure}, or \ref{U1Mr023b001figure} and \ref{U1Mr023b002figure}). It is also possible to obtain a solution where both the $\mathrm{FM}$ and $\mathrm{AFM}$ order parameters are finite and comparable, on top of which the spin liquid parameters may be zero or finite, which we label by $\mathrm{M}$ or $\mathrm{M}^*$. The $\mathrm{M}$ and $\mathrm{M}^*$ phases are absent in the zero field limit. When the $\mathrm{FM}$ and $\mathrm{AFM}$ order parameter have about the same magnitude (e.g. $\lvert \mathbf{m}_{\mathrm{T}_{1,\mathrm{A}'}} \rvert \sim 0.5$ and $\lvert \mathbf{m}_\mathrm{E} \rvert \sim 0.5$), the spin configuration of the M phase can only be known by calculating the expectation value of spin operators at the four sublattices of the pyrochlore unit cell. Otherwise, if one of the $\mathrm{FM}$ and $\mathrm{AFM}$ order parameters is much larger than the other (e.g. $\lvert \mathbf{m}_{\mathrm{T}_{1,\mathrm{A}'}} \rvert \sim 0.9$ and $\lvert \mathbf{m}_\mathrm{E} \rvert \sim 0.1$), then the spin configuration of the $\mathrm{M}$ phase will of course resemble the dominant order. Eventually, when the field strength is sufficiently large, the phase region with deconfined spinons vanishes entirely and the system becomes classical in the neighborhood of Gaulin and Coldea parametrizations (see Fig.~\ref{Z2Ur023b004figure} or \ref{U1Mr023b004figure}).

\subsection{\label{localglobalminima}Local and Global Minima}

We pick a representative value of the weighting factor $r=0.23$ to extract some qualitative features of the mean field solutions at the Gaulin and Coldea parametrizations. In the zero field limit, with the $\mathbb{Z}_2\mathrm{U}$ ansatz, these parametrizations are in the fractionalized magnetically ordered phases, but very close to the pure spin liquid phase (see Fig.~\ref{Z2Ur023b000figure}). On the other hand, with the  $U(1)\mathrm{M}$ ansatz, the magnetic order is very weak (i.e.~$S/S_0$ is very small) in the $\mathrm{FM}^*$ and $\mathrm{AFM}^*$ phases, at Gaulin and Coldea parametrizations respectively (see Fig.~\ref{U1Mr023b000figure} and Tables \ref{U1Mgaulintable} and \ref{U1Mcoldeatable} in Appendix \ref{localglobalcomparison}). These suggest that a pure spin liquid phase is energetically competitive with the fractionalized magnetically ordered ground states.
%We pick a representative value of the weighting factor $r=0.23$ to extract some qualitative features of the mean field solutions for Gaulin and Coldea parametrizations. First, in the zero field limit, when we consider the $\mathbb{Z}_2\mathrm{U}$ ansatz,  both of Gaulin and Coldea parametrizations are in a fractionalized magnetically ordered phase, but very close to the pure spin liquid state (see Fig.~\ref{Z2Ur023b000figure}). Secondly, when we consider the $U(1)\mathrm{M}$ ansatz, both of Gaulin and Coldea parametrizations are again in a fractionalized magnetically ordered phase while there is no nearby pure spin liquid stare in the phase diagram. However, these fractionalized magnetically ordered states have very small magnetic order parameters. Hence in all of these cases, it is clear that the pure spin liquid states are energetically competitive in comparison to the fractionalized magnetically ordered ground states.
Indeed, we find that the pure spin liquid state is another convergent solution from the self consistent calculations, but corresponds to a local minimum, for the $\mathbb{Z}_2\mathrm{U}$ ansatz. We compare the energies of the two mean field solutions corresponding to the local and global minima, where the spin liquid ($S/S_0 \ll 1$) and magnetic order ($S/S_0 \sim 1$) dominate respectively, in Tables \ref{Z2Ugaulintable} and \ref{Z2Ucoldeatable} in Appendix \ref{localglobalcomparison}. Similar tables are constructed for the $U(1)\mathrm{M}$ ansatz, where the spin liquid dominant coexisting state is also the ground state at small enough fields (see Table \ref{U1Mgaulintable} and \ref{U1Mcoldeatable} in Appendix \ref{localglobalcomparison}). As the magnetic field $B_z$ increases in strength, the $\mathrm{FM}$ state is more favorable, and the difference between the local and global minima becomes more significant. When the field strength is sufficiently large ($B_z \gtrsim 0.01 \vert J_3 \vert$), we can no longer get the (spin liquid dominant) $\mathrm{FM}^*$ phase, the self consistent calculation always yields the $\mathrm{FM}$ phase, and the system becomes fully classical. The coexisting solution for the Coldea parametrization is relatively more persistent with increasing field compared to the Gaulin parametrization.

\subsection{\label{dispersionstructurefactor}Spinon Dispsersion and Dynamical Spin Structure Factor}

\begin{figure}
\includegraphics[scale=0.4]{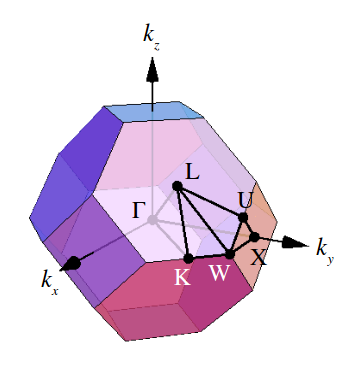}
\caption{\label{brillouinfigure}The first Brillouin zone of the fcc lattice (the underlying Bravais lattice of the pyrochlore lattice). Several points of high symmetry are indicated.}
\end{figure}

\begin{figure}
\subfloat[]{\label{Z2Ugaulinbandr023b0000sl} \includegraphics[scale=0.24]{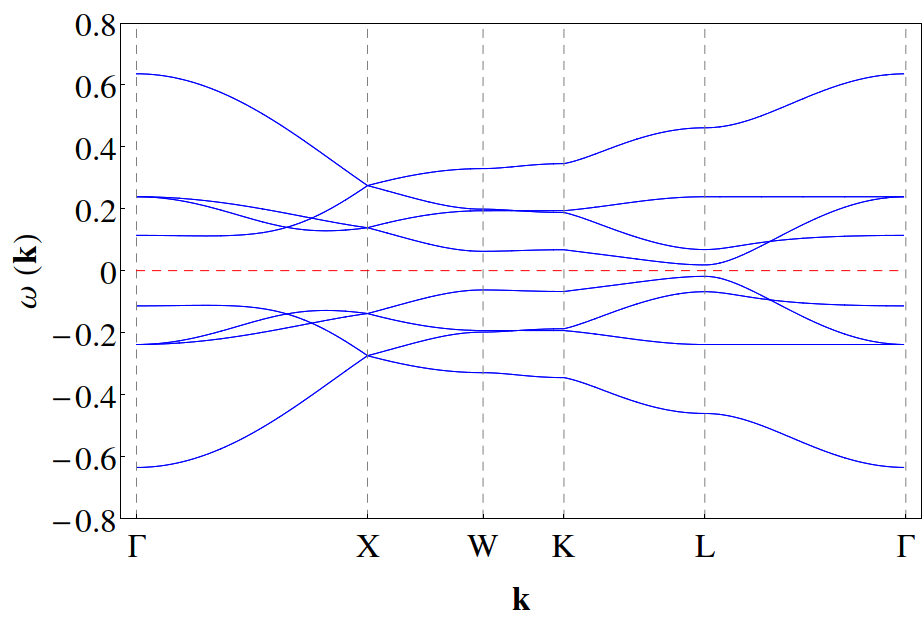}} \\
\subfloat[]{\label{Z2Ugaulinbandr023b0010fmsl} \includegraphics[scale=0.24]{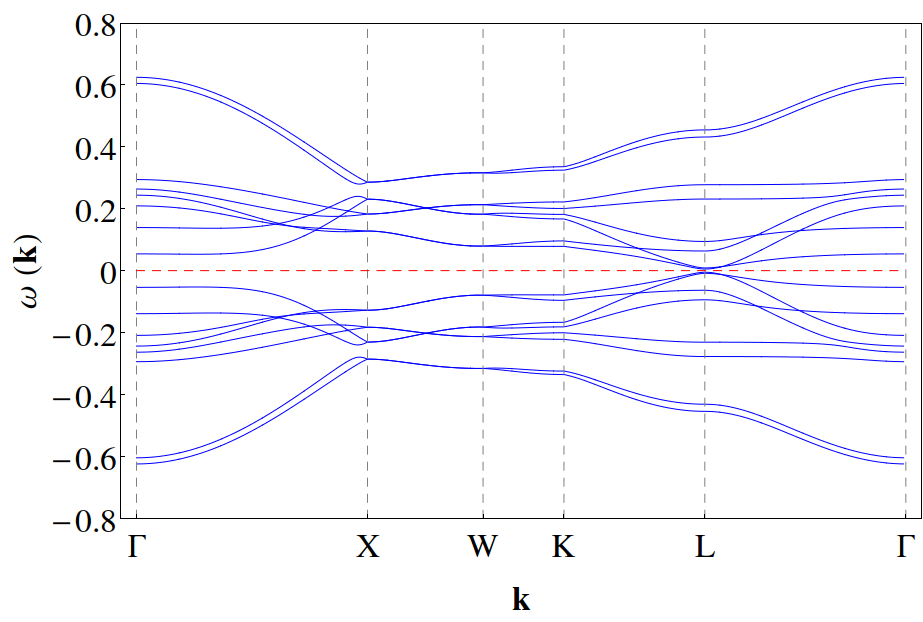}}
\caption{(a) The spinon band structure in the $\mathbb{Z}_2 \mathrm{U}$ spin liquid phase, which is a local minimum of $H^\mathrm{MF}$, at zero magnetic field. Each band is two fold degenerate because of the presence of both inversion and time reversal symmetries. The red dashed horizontal line indicates the zero level, above which the excitation spectrum of Bogoliubov quasiparticles of spinons lies. (b) A small but finite magnetic field, e.g. $B_z/\lvert J_3 \rvert=0.01$ as shown here, turns on the $\mathrm{FM}$ order parameter and lifts the two fold degeneracy. The resulting spin liquid dominant $\mathrm{FM}^*$ phase is still a local minimum. See Table \ref{Z2Ugaulintable}.}
\end{figure}

\begin{figure}
\subfloat[]{\label{U1Mgaulinbandr023b0000sl} \includegraphics[scale=0.24]{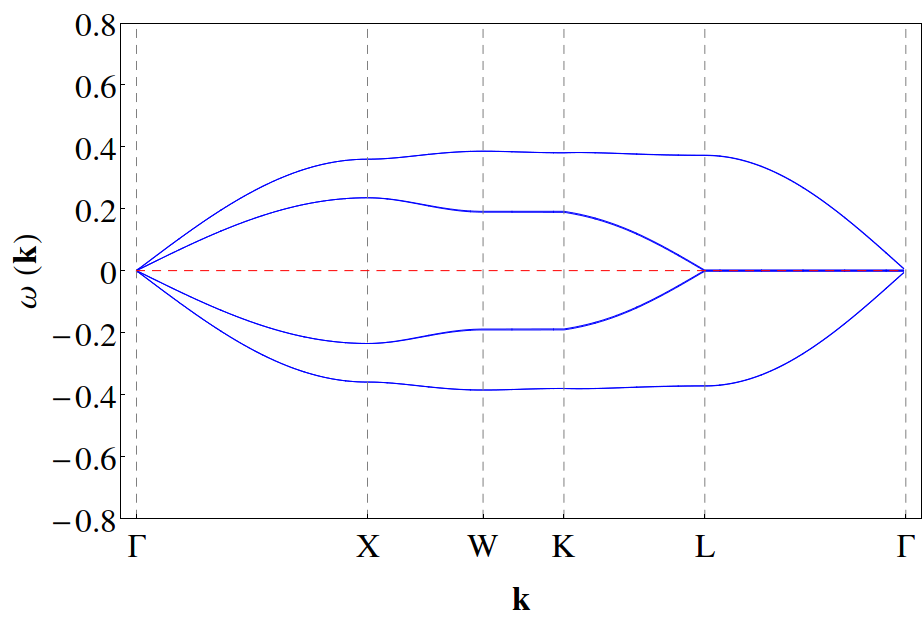}} \\
\subfloat[]{\label{U1Mgaulinbandr023b0010fmsl} \includegraphics[scale=0.24]{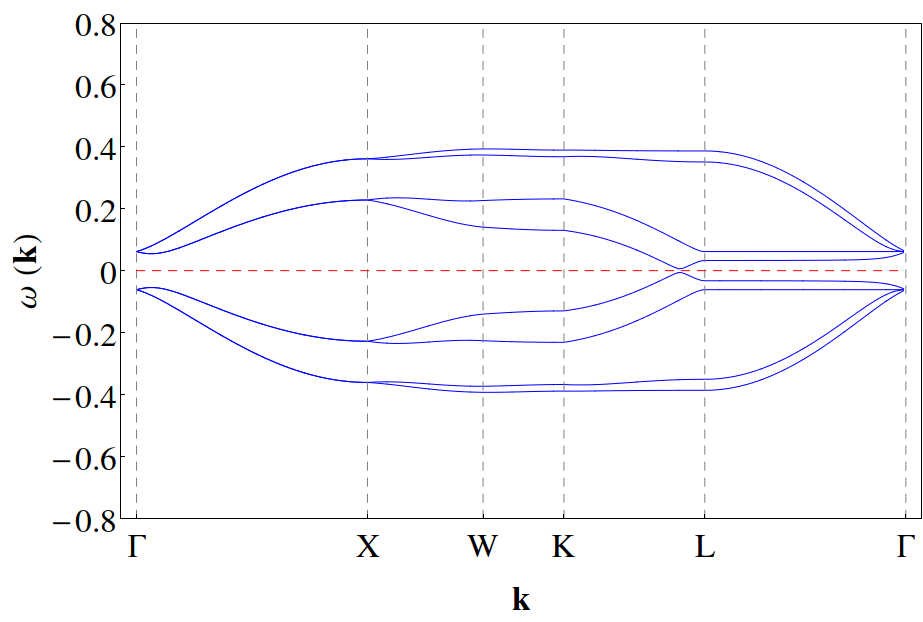}}
\caption{(a) The spinon band structure in the $U(1) \mathrm{M}$ spin liquid phase at zero magnetic field. Each band is two fold degenerate as the combination of inversion and time reversal is a symmetry. The red dashed line indicates the Fermi level of spinons, which is determined by the single occupancy constraint \eqref{singleoccupancy}. (b) A small but finite magnetic field, e.g. $B_z/\lvert J_3 \rvert=0.01$ as shown here, lifts the two fold degeneracy. The FM order parameter becomes nonnegligible (see Table \ref{U1Mgaulintable}), and the resulting spin liquid dominant $\mathrm{FM}^*$ phase is a local minimum of $H^\mathrm{MF}$.}
\end{figure}

\begin{figure}
\subfloat[]{\label{Z2Ugaulinbandr023b0000gs} \includegraphics[scale=0.24]{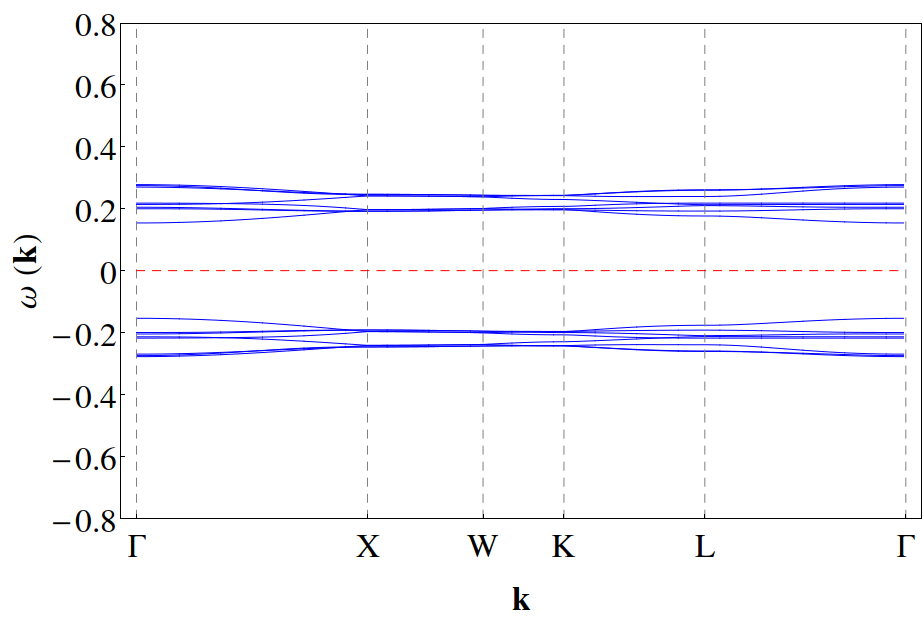}} \\
\subfloat[]{\label{Z2Ucoldeabandr023b0000gs} \includegraphics[scale=0.24]{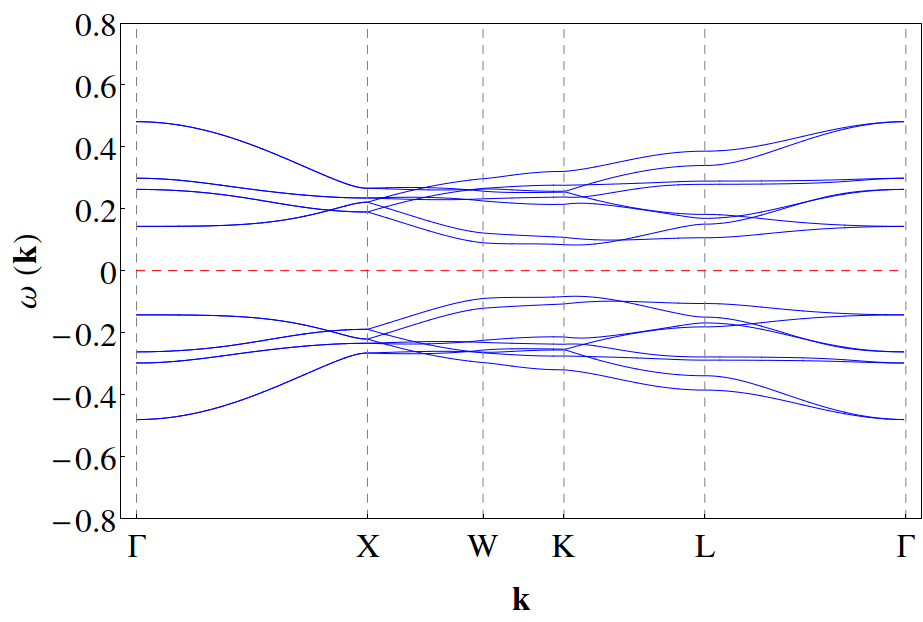}} \\
\subfloat[]{\label{Z2Ugaulinbandr023b0004gs} \includegraphics[scale=0.24]{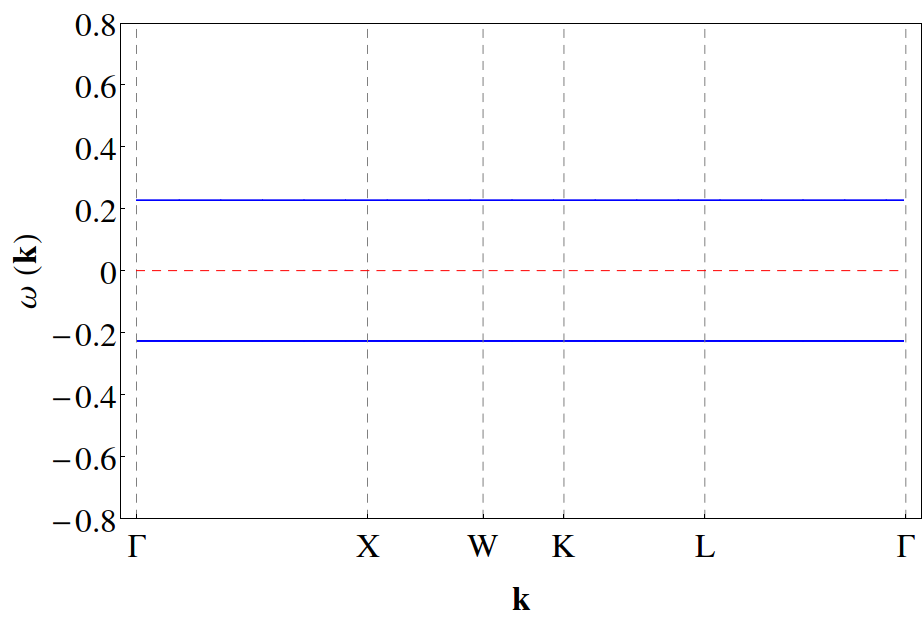}}
\caption{The spinon band structure in the ground state of $H^\mathrm{MF}$ with the $\mathbb{Z}_2 \mathrm{U}$ ansatz at zero magnetic field at (a) Gaulin and (b) Coldea parametrizations. The ground state is a magnetic order dominant $\mathrm{FM}^*$ or $\mathrm{AFM}^*$ phase, so that the spinon dispersion is relatively flat and the excitation gap is relatively large compared to that in Fig.~\ref{Z2Ugaulinbandr023b0000sl} or \ref{U1Mgaulinbandr023b0000sl}. (c) At Gaulin parametrization, the spinon dispersion is completely flattened out at $B_z=0.004 \lvert J_3 \rvert$ when the system enters the pure $\mathrm{FM}$ phase. Each band is four fold degenerate.}
\end{figure}

\begin{figure}
\subfloat[]{\label{Z2Uintensitygaulinr023b0000gs} \includegraphics[scale=0.6]{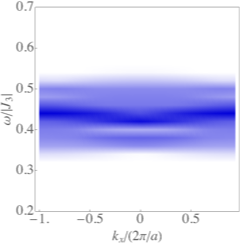}} \\
\subfloat[]{\label{Z2Uintensitygaulinr023b0000sl} \includegraphics[scale=0.6]{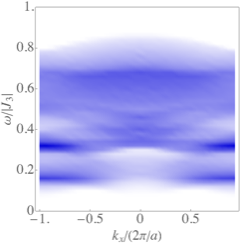}} \\
\subfloat[]{\label{U1Mintensitygaulinr023b0000gs} \includegraphics[scale=0.6]{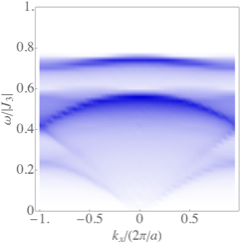}}
\caption{The dynamical spin structure factor at Gaulin parametrization in the zero field limit: (a) the magnetic-order-dominant $\mathrm{FM}^*$ ground state with the $\mathbb{Z}_2 \mathrm{U}$ ansatz, (b) the pure spin liquid state with the $\mathbb{Z}_2 \mathrm{U}$ ansatz, and (c) the spin-liquid-dominant $\mathrm{FM}^*$ ground state with the $U(1)\mathrm{M}$ ansatz. Darker regions indicate higher intensities. A broad (narrow) continuum is obtained when the spin liquid (magnetic order) parameters dominate. Notice that the dynamical spin structure factor of the $\mathrm{FM}^*$ ground state obtained with the $U(1)\mathrm{M}$ ansatz is very similar to that of the pure $U(1)\mathrm{M}$ spin liquid state (not shown) because the magnetic order is very weak.}
\end{figure}

We examine the spinon band structures of the pure spin liquid (or spin liquid dominant) phases with the $\mathbb{Z}_2\mathrm{U}$ and $U(1)\mathrm{M}$ ansatzes along some high symmetry directions in the Brillouin zone (see Fig.~\ref{brillouinfigure}).\cite{SETYAWAN2010299} As Gaulin and Coldea parametrizations are considerably close in phase space (see Fig.~\ref{classicalphasediagramfigure}), the spinon dispersions of the pure spin liquid (or spin liquid dominant) phases for these parametrizations are quite similar. For simplicity, we only show the spinon dispersion at Gaulin parametrization. At zero field, the $\mathbb{Z}_2\mathrm{U}$ spin liquid has a small gap (of the order of $0.01 \lvert J3 \rvert$), while the $U(1)\mathrm{M}$ spin liquid is gapless (see Fig.~\ref{Z2Ugaulinbandr023b0000sl} and \ref{U1Mgaulinbandr023b0000sl}). The bands are two fold degenerate in both cases as, for the $\mathbb{Z}_2\mathrm{U}$ ansatz, the inversion and time reserval symmetries are present, while for the $U(1)\mathrm{M}$ ansatz, although the inversion and time reserval symmetries are broken separately, the combination of them is a symmetry.\cite{PhysRevB.79.144432} The degeneracy is lifted at finite fields (see Figs.~\ref{Z2Ugaulinbandr023b0010fmsl} and \ref{U1Mgaulinbandr023b0010fmsl}), when the magnetic order parameter becomes significant.

We also look at some instances of the spinon band structures in the magnetic order dominant coexisting phases, say, with the $\mathbb{Z}_2\mathrm{U}$ ansatz. At zero field, the Gaulin and Coldea parametrizations falls into $\mathrm{FM}^*$ and $\mathrm{AFM}^*$ phases respectively (see Fig.~\ref{Z2Ur023b000figure}), so their dispersions do not resemble each other (see Figs.~\ref{Z2Ugaulinbandr023b0000gs} and \ref{Z2Ucoldeabandr023b0000gs}). The spinon dispersion is relatively flat, and the excitation gap is relatively large, compared to those in the spin liquid dominant coexisting phases. Under the magnetic field, the magnetic order parameter (spin liquid parameters) further increases (decrease), the bands becomes less dispersing and eventually completely flatten out in the purely magnetic ($\mathrm{FM}$) phase when the field is sufficiently large (see Fig.~\ref{Z2Ugaulinbandr023b0004gs}). The flat bands are four fold degenerate (see Appendix \ref{magneticband} for explanation).

Either in the spin liquid or magnetic order dominant coexisting phases, as long as the spin liquid parameters are not all zero, we will get dispersing spinon bands and thus, a two spinon continuum, which is related to the dynamical spin structure factor,
\begin{equation}
S(\mathbf{k},\omega) = \sum_{ij} e^{-i \mathbf{k} \cdot \left( \mathbf{R}_i - \mathbf{R}_j \right) } \int \mathrm{d} t e^{i \omega t} \langle \mathbf{S}_i \left( t \right) \cdot \mathbf{S}_j \left( 0 \right) \rangle
\end{equation}
We calculate the dynamical spin structure factor for a few illustrative cases along the $k_x$ direction. The width of the continuum depends on the relative weight of the spin liquid parameters to the magnetic order parameters. If the spin liquid parameters dominate over the magnetic order parameters, meaning that the ratio $S/S_0 \longrightarrow 0$ is small and the quantum effect is strong, then a broad continuum is obtained (see Fig.~\ref{Z2Uintensitygaulinr023b0000sl} and \ref{U1Mintensitygaulinr023b0000gs}). If the converse is true, then a narrow continuum is obtained (see Fig.~\ref{Z2Uintensitygaulinr023b0000gs}).

\section{\label{discussion}Summary and Discussion}

Recent inelastic neutron scattering experiments\cite{PhysRevX.1.021002, PhysRevLett.119.057203} on $\mathrm{Yb}_2\mathrm{Ti}_2\mathrm{O}_7$ put stringent constraints on possible spin models and suggest that the system is very close to the classical phase boundary between a splayed ferromagnet and an antiferromagnet. While the ground state is in the splayed ferromagnetic phase, Ref.~\onlinecite{PhysRevLett.119.057203} finds that the spin excitation spectrum is characterized by a continuum and spin wave excitations seem to break down in the absence of external magnetic field. In this work, we investigate the possibility of stabilizing a magnetically ordered phase with deconfined spinon excitations (fractionalized magnetically ordered phase) in the vicinity of the classical phase boundary mentioned above, using the spin models given by two different sets of exchange parameters, dubbed Gaulin\cite{PhysRevX.1.021002} and Coldea\cite{PhysRevLett.119.057203} parametrizations.

The generic spin model with nearest neighbor exchange interactions on the pyrochlore lattice \eqref{spinhamiltonian} contains four exchange parameters $J_1$, $J_2$, $J_3$, and $J_4$\cite{PhysRevX.1.021002,PhysRevB.95.094422} in the global coordinate, but we set $J_4=0$ for simplicity as it is one order of magnitude smaller than $J_2, J_3$ in both Gaulin and Coldea parametrizations. With this simplification, the spin Hamiltonian in the conventional basis has the form of the $JK\Gamma$ model. Gaulin and Coldea parametrizations suggest that the highly frustrating interactions, $K$ and $\Gamma$, are large in $\mathrm{Yb}_2\mathrm{Ti}_2\mathrm{O}_7$.

We consider the complex fermion mean field theory of the $JK\Gamma$ model that includes both the spin liquid and magnetic order channels on equal footing. We use the $\mathbb{Z}_2$ uniform ansatz and $U(1)$ monopole flux ansatz\cite{PhysRevB.79.144432} in the spin liquid part of the Hamiltonian. For the magnetic part, we take into account the competing splayed ferromagnetic and antiferromagnetic orders. We introduce the relative weighting factor, $r \in [0,1]$, for the spin liquid and the magnetic order in the mean field theory, with smaller $r$ corresponding to stronger quantum fluctuations. With intermediate values of $r$, we find that the magnetically ordered phase with deconfined spinons is a stable mean field ground state, where the magnetic order parameter is reduced from its classical value due to quantum fluctuations represented by spinon excitations. We then compute the dynamical spin structure factor, which shows a scattering continuum arising from two spinon excitations, as was observed in the inelastic neutron scattering experiment at zero magnetic field. Upon increasing the field, quantum fluctuations are suppressed and a conventional magnetic order with no fractionalized excitations become the ground state, which is consistent with the experimental finding that the high field splayed ferromagnetic phase has well defined spin wave excitations. 

In this work, we map out the phase diagram by varying the weighting factor $r$, which arises from the ambiguity in writing down the total mean field Hamiltonian. As mentioned earlier, in principle, $r$ may be determined dynamically if there is a way to go beyond 
the mean field theory. At present, there is no systematic way to determine which value of $r$ should be chosen within a mean field theory. On the other hand, we believe that some intermediate values of $r$ may correspond to the physical limit as significant quantum fluctuations must be present in the quantum ground state of the spin model corresponding to Gaulin and Coldea parametrizations. It would be great if there is a way to estimate the appropriate value of $r$ with an analysis similar to the application of the Gutzwiller approximation/projection in the $tJ$ model, which leads to renormalization of the hopping 
integral $t$ and the Heisenberg interaction $J$.\cite{cond-mat/0003465}

We consider here only two quantum spin liquid ansatzes, the $\mathbb{Z}_2$ uniform and $U(1)$ monopole flux states, which are allowed by the projective symmetry group (PSG) of the pyrochlore lattice. Certainly there are many other competing spin liquid states that may also permit a coexisting magnetic order. In order to carry out a more systematic investigation, one will have to classify all the possible fermionic spin liquid states on the pyrochlore lattice. Future work on this issue will be desirable for a more
complete analysis of the possible spin liquid and fractionalized magnetically ordered phases in $\mathrm{Yb}_2\mathrm{Ti}_2\mathrm{O}_7$.

\begin{acknowledgments}
We are grateful to Wonjune Choi and Arun Paramekanti for useful discussions. We also thank Chunxiao Liu and Leon Balents for explaining their work on the classification of bosonic spin liquid states on the pyrochlore lattice, which we did not consider in our study. This work was supported by the NSERC of Canada and the Center for Quantum Materials at the University of Toronto. Most of the computations were performed on the GPC supercomputer at the SciNet HPC Consortium.\cite{1742-6596-256-1-012026} SciNet is funded by: the Canada Foundation for Innovation under the auspices of Compute Canada; the Government of Ontario; Ontario Research Fund - Research Excellence; and the University of Toronto.
\end{acknowledgments}

\appendix

\section{\label{localcoordinates}Local Coordinates, Spin Hamiltonian, and $g$ Tensor}

The bases in the local coordinates of the four sublattices are defined as\cite{PhysRevX.1.021002}
\begin{subequations}
\begin{align}
& \hat{\mathsf{z}}_0=\frac{1}{\sqrt{3}} \left(1,1,1\right), \hat{\mathsf{x}}_0 = \frac{1}{\sqrt{6}} \left(-2,1,1\right); \label{localbases0} \\
& \hat{\mathsf{z}}_1=\frac{1}{\sqrt{3}} \left(1,-1,-1\right), \hat{\mathsf{x}}_1 = \frac{1}{\sqrt{6}} \left(-2,-1,-1\right); \label{localbases1} \\
& \hat{\mathsf{z}}_2=\frac{1}{\sqrt{3}} \left(-1,1,-1\right), \hat{\mathsf{x}}_2 = \frac{1}{\sqrt{6}} \left(2,1,-1\right); \label{localbases2} \\
& \hat{\mathsf{z}}_3=\frac{1}{\sqrt{3}} \left(-1,-1,1\right), \hat{\mathsf{x}}_3 = \frac{1}{\sqrt{6}} \left(2,-1,1\right). \label{localbases3}
\end{align}
\end{subequations}

The unimodular complex numbers in the local Hamiltonian \eqref{spinhamiltonianlocal} are given by\cite{PhysRevX.1.021002}
\begin{subequations}
\begin{align}
\zeta &= \begin{pmatrix}
0 & -1 & e^{i \pi /3} & e^{-i \pi /3} \\
-1 & 0 & e^{-i \pi /3} & e^{i \pi /3} \\
e^{i \pi /3} & e^{-i \pi /3} & 0 & -1 \\
e^{-i \pi /3} & e^{i \pi /3} & -1 & 1 \\
\end{pmatrix} , \label{zetamatrix} \\
\gamma &= -\zeta^*. \label{gammamatrix}
\end{align}
\end{subequations}

To obtain the global exchange parameters in \eqref{spinhamiltonian} from the local exchange parameters in \eqref{spinhamiltonianlocal}, we just have to rotate the local bases $(\hat{\mathsf{x}}_s,\hat{\mathsf{y}}_s,\hat{\mathsf{z}}_s)$ such that they align with the global bases $(\hat{\mathbf{x}},\hat{\mathbf{y}},\hat{\mathbf{z}})$. Call these sublattice dependent rotations $\mathsf{R}_s$. We then have, for example,
\begin{equation} \label{globallocalexchangeexample}
\mathsf{J}_{01}^\mathrm{global}=\mathsf{R}_0^{-1} \mathsf{J}_{01}^\mathrm{local} \mathsf{R}_1 .
\end{equation}
The final result is
\begin{equation} \label{globallocalexchangerelation}
\begin{pmatrix} J_1 \\ J_2 \\ J_3 \\ J_4 \end{pmatrix} = \frac{1}{3} \begin{pmatrix} -1 & 4 & 2 & 2 \sqrt{2} \\ 1 & -4 & 4 & 4 \sqrt{2} \\ -1 & -2 & -4 & 2 \sqrt{2} \\ -1 & -2 & 2 & -\sqrt{2} \end{pmatrix} \begin{pmatrix} J_{zz} \\ J_{\pm} \\ J_{\pm \pm} \\ J_{z \pm} \end{pmatrix} .
\end{equation}

To relate the interactions on different bonds, in the global coordinates, we can use the $C_3$ rotations, for instance
\begin{equation} \label{globallocalhamiltonianexample}
\mathsf{J}_{02}^\mathrm{global} = O_{C_3^{[111]}} \mathsf{J}_{01}^\mathrm{global} O_{C_3^{[111]}}^{-1} .
\end{equation}
The expression of $O_{C_3^{[111]}}$ can be found in \eqref{so3matrix}. We list all these interactions below for completeness.\cite{PhysRevB.95.094422}
\begin{equation} \label{spinhamiltonianglobal}
\begin{aligned}[b]
& \mathsf{J}_{01}^\mathrm{global} = \begin{pmatrix} J_2 & J_4 & J_4 \\ -J_4 & J_1 & J_3 \\ -J_4 & J_3 & J_1 \end{pmatrix}, \, \mathsf{J}_{02}^\mathrm{global} = \begin{pmatrix} J_1 & -J_4 & J_3 \\ J_4 & J_2 & J_4 \\ J_3 & -J_4 & J_1 \end{pmatrix} , \\
& \mathsf{J}_{03}^\mathrm{global} = \begin{pmatrix} J_1 & J_3 & -J_4 \\ J_3 & J_1 & -J_4 \\ J_4 & J_4 & J_2 \end{pmatrix}, \, \mathsf{J}_{12}^\mathrm{global} = \begin{pmatrix} J_1 & -J_3 & J_4 \\ -J_3 & J_1 & -J_4 \\ -J_4 & J_4 & J_2 \end{pmatrix} , \\
& \mathsf{J}_{23}^\mathrm{global} = \begin{pmatrix} J_2 & -J_4 & J_4 \\ J_4 & J_1 & -J_3 \\ -J_4 & -J_3 & J_1 \end{pmatrix}, \, \mathsf{J}_{31}^\mathrm{global} = \begin{pmatrix} J_1 & -J_4 & -J_3 \\ J_4 & J_2 & -J_4 \\ -J_3 & J_4 & J_1 \end{pmatrix} .
\end{aligned}
\end{equation}

The $g$ tensor in global coordinates, which is sublattice dependent, can be obtained from that in local coordinates by rotations of the bases similar to the consideration in \eqref{globallocalexchangeexample}. That is,
\begin{equation}
\mathsf{g}_s^\mathrm{global} = \mathsf{R}_s^{-1} \mathsf{g}^\mathrm{local} \mathsf{R}_s
\end{equation}
We list all the $g$ tensors below for completeness.\cite{PhysRevB.95.094422}
\begin{equation} \label{gtensorglobal}
\begin{aligned}[b]
& \mathsf{g}_{s=0}^\mathrm{global} = \begin{pmatrix} g_1 & g_2 & g_2 \\ g_2 & g_1 & g_2 \\ g_2 & g_2 & g_1 \end{pmatrix}, \, \mathsf{g}_{s=1}^\mathrm{global} = \begin{pmatrix} g_1 & -g_2 & -g_2 \\ -g_2 & g_1 & g_2 \\ -g_2 & g_2 & g_1 \end{pmatrix}, \\
& \mathsf{g}_{s=2}^\mathrm{global} = \begin{pmatrix} g_1 & -g_2 & g_2 \\ -g_2 & g_1 & -g_2 \\ g_2 & -g_2 & g_1 \end{pmatrix}, \, \mathsf{g}_{s=3}^\mathrm{global} = \begin{pmatrix} g_1 & g_2 & -g_2 \\ g_2 & g_1 & -g_2 \\ -g_2 & -g_2 & g_1 \end{pmatrix},
\end{aligned}
\end{equation}
where $g_1=2g_{xy}/3+g_z/3$ and $g_2=-g_{xy}/3+g_z/3$.

\section{\label{spinliquidansatzesdetails}Details of the Spin Liquid Ansatzes}

We discuss the $\mathbb{Z}_2 \mathrm{U}$ and $U(1) \mathrm{M}$ spin liquid ansatzes in details, especially the interdependence of the spinon hopping and pairing parameters in $H_\mathrm{SL}^\mathrm{MF}$. The allowed forms of these mean field parameters are dictated by the symmetries of the system. Constraint arises when one symmetry element maps a bond to itself, or two different symmetry elements relates two different bonds.

\subsection{\label{Z2Uansatzdetails} $\mathbb{Z}_2$ Uniform Ansatz}

We first introduce the following $2 \times 2$ matrix whose components are the spinon creation and annihilation operators,\cite{PhysRevB.95.054404}
\begin{equation} \label{spinonmatrix}
\Psi_i = \begin{pmatrix} f_{i \uparrow} & f_{i \downarrow} \\ f_{i \downarrow}^\dagger & - f_{i \uparrow}^\dagger \end{pmatrix} .
\end{equation}
The spin operator \eqref{secondquantizationspinoperator} can then be expressed as
\begin{equation} \label{spinoperatormatrix}
S^\mu_i = \frac{1}{4} \mathrm{Tr} \left( \Psi_i^\dagger \sigma^\mu \Psi_i \right),
\end{equation}
and the spin liquid Hamiltonian \eqref{HSL} after the mean field decoupling \eqref{meanfielddecoupling} as
\begin{equation} \label{HSLMFmatrix}
H_\mathrm{SL}^\mathrm{MF} = \sum_{ij} \sum_{\mu=0,x,y,z} \mathrm{Tr} \left( \sigma^\mu \Psi_i u_{ij}^\mu \Psi_j^\dagger \right) ,
\end{equation}
where $u_{ij}^\mu$ are $2 \times 2$ matrices of the mean field ansatzes. For instance, on the bond $\langle 01 \rangle$, with the exchange couplings $J,K,\Gamma < 0$,
\begin{subequations}
\begin{align*}
u_{01}^0 &= \frac{\lvert \Gamma \rvert}{8} \begin{pmatrix} \chi_{01} & - \Delta_{01}^* \\ - \Delta_{01} & - \chi_{01}^* \end{pmatrix} \\
u_{01}^x &= \frac{2 \lvert J \rvert + \lvert \Gamma \rvert}{8} \begin{pmatrix} E_{01}^x & D_{01}^{x*} \\ - D_{01}^x & E_{01}^{x*}\end{pmatrix} \\
u_{01}^y &= \frac{2 \lvert J \rvert + \lvert K \rvert + \lvert \Gamma \rvert}{8} \begin{pmatrix} E_{01}^y & D_{01}^{y*} \\ - D_{01}^y & E_{01}^{y*} \end{pmatrix} - \frac{\lvert \Gamma \rvert}{8} \begin{pmatrix} E_{01}^z & D_{01}^{z*} \\ - D_{01}^{z} & E_{01}^{z*} \end{pmatrix} \\
u_{01}^z &= \frac{2 \lvert J \rvert + \lvert K \rvert + \lvert \Gamma \rvert}{8} \begin{pmatrix} E_{01}^z & D_{01}^{z*} \\ - D_{01}^z & E_{01}^{z*} \end{pmatrix} - \frac{\lvert \Gamma \rvert}{8} \begin{pmatrix} E_{01}^y & D_{01}^{y*} \\ - D_{01}^{y} & E_{01}^{y*} \end{pmatrix}
\end{align*}
\end{subequations}
We also have
\begin{equation}
u_{ii}^0 = \begin{pmatrix} \mu_3 & \mu_1 - i \mu_2 \\ \mu_1 + i \mu_2 & - \mu_3 \end{pmatrix} \label{singleoccupancymatrix}
\end{equation}
that enforces the single occupancy constraint \eqref{singleoccupancy}.
In the form \eqref{spinoperatormatrix}, it is now obvious that the spinon representation of spin is invariant under an $SU(2)$ gauge transformation
\begin{equation} \label{gaugetransformation}
\Psi_i \longrightarrow \Psi_i G_i, \, G_i \in SU(2) ,
\end{equation}
which has been mentioned in Section \ref{spinliquidansatzes}. We apply the symmetry operations passively, that is, transform the coordinate axes forward (equivalently transform the vectors backward),\cite{PhysRevB.94.035107} such that
\begin{equation}
\mathbf{S}_i \overset{X}{\longrightarrow} R_X^{-1} \mathbf{S}_{X(i)} , \label{transformspin}
\end{equation}
where $X$ is an element of the space group and $R_X$ is the $SU(2)$ spin rotation associated with $X$. In the representation \eqref{spinoperatormatrix}, the symmetry transformation \eqref{transformspin} is achieved by\cite{PhysRevB.95.054404}
\begin{equation}
\Psi_i \overset{X}{\longrightarrow} e^{i \sigma \cdot \hat{\mathbf{n}} \phi/2}\Psi_{X(i)} , \label{transformspinonmatrix}
\end{equation}
where $\hat{\mathbf{n}}$ is a unit vector along the axis of rotation and $\phi$ is the angle of rotation associated with $X$. Therefore, $X$ acts on the mean field Hamiltonian \eqref{HSLMFmatrix} by
\begin{equation}
\begin{aligned}[b]
H_\mathrm{SL}^\mathrm{MF} & \overset{X}{\longrightarrow} \sum_{ij} \sum_{\mu=0,x,y,z} \mathrm{Tr} \left( e^{-i \sigma \cdot \hat{\mathbf{n}} \phi/2} \sigma^\mu e^{i \sigma \cdot \hat{\mathbf{n}} \phi/2} \Psi_{X(i)} u_{ij}^\mu \Psi_{X(j)}^\dagger \right) \\ 
&= \sum_{ij} \mathrm{Tr} \left( \Psi_{X(i)} u_{ij}^0 \Psi_{X(j)}^\dagger \right) \\
& \quad + \sum_{ij} \sum_{\mu=x,y,z} \mathrm{Tr} \left( \sum_{\nu=x,y,z} O_X^{-1 \, \mu \nu} \sigma^\nu \Psi_{X(i)} u_{ij}^\mu \Psi_{X(j)}^\dagger \right) \label{transformHSLMFmatrix}
\end{aligned}
\end{equation}
where the $SU(2)$ spin rotation $R_X$ has been mapped to the $SO(3)$ rotation $O_X$ on the Pauli matrices. The Hamiltonian should be left invariant under $X$ by the definition of symmetry. Taking into account the $SU(2)$ gauge redundancy \eqref{gaugetransformation}, this implies that the mean field ansatzes should obey the relations
\begin{subequations}
\begin{align}
u_{X(i)X(j)}^0 &= G_X \left( X(i) \right) u_{ij}^0 G_X \left( X(j) \right)^\dagger \label{singletspacepsg} \\
u_{X(i)X(j)}^{\mu=x,y,z} &= \sum_{\nu=x,y,z} O_X^{\mu \nu} G_X \left( X(i) \right) u_{ij}^\nu G_X \left( X(j) \right)^\dagger \label{tripletspacepsg}
\end{align}
\end{subequations}
where $G_X \left( i \right)$ is the $SU(2)$ gauge transformation associated with $X$ at site $i$. To this end, we list the $SO(3)$ matrices $O_X$ associated with some representative elements of the $\mathrm{Fd}\bar{3}\mathrm{m}$ space group discussed in Section \ref{structuresymmetry},
\begin{equation} \label{so3matrix}
\begin{aligned}[b]
& O_{C_3^{[111]}} = \begin{pmatrix} 0 & 0 & 1 \\ 1 & 0 & 0 \\ 0 & 1 & 0 \end{pmatrix} , \,
O_{C_2^{x}} = \begin{pmatrix} 1 & 0 & 0 \\ 0 & -1 & 0 \\ 0 & 0 & -1 \end{pmatrix} , \,
O_{S_4^{x}} = \begin{pmatrix} 1 & 0 & 0 \\ 0 & 0 & -1 \\ 0 & 1 & 0 \end{pmatrix} , \\
& O_{\sigma_\mathrm{d}^{[011]}} = \begin{pmatrix} -1 & 0 & 0 \\ 0 & 0 & 1 \\ 0  & 1 & 0 \end{pmatrix} , \,
O_\mathcal{I} = \begin{pmatrix} 1 & 0 & 0 \\ 0 & 1 & 0 \\ 0 & 0 & 1 \end{pmatrix} .
\end{aligned}
\end{equation}
All other space group elements can be constructed from these, e.g. $C_2^y = C_3^{[111]} C_2^x C_3^{[111] -1}$. The two fold rotation itself can be obtained by twice the four fold improper rotations, e.g. $C_2^x = (S_4^x)^2$. Note that, since spin is a pseudovector, it is invariant under inversion, hence the $SO(3)$ matrix associated with inversion is the identity. The reflections and improper rotations can be viewed as a combination of rotation and inversion, and their corresponding $SO(3)$ matrices only encode the rotation. For example, the reflection $\sigma_\mathrm{d}$ across the plane perpendicular to the $[011]$ direction is a rotation by $\pi$ about the $[011]$ axis followed by inversion about the intersection of the axis and the plane.

On the other hand, time reversal $\mathcal{T}$ acts on the mean field Hamiltonian \eqref{HSLMFmatrix} by
\begin{equation}
\begin{aligned}[b]
H_\mathrm{SL}^\mathrm{MF} & \overset{\mathcal{T}}{\longrightarrow} \sum_{ij} \sum_{\mu=0,x,y,z} \mathrm{Tr} \left(-i \sigma^y \sigma^{\mu *} i \sigma^y \Psi_i u_{ij}^{\mu *} \Psi_j^\dagger \right) \\
&= \sum_{ij} \mathrm{Tr} \left(\Psi_i u_{ij}^{0 *} \Psi_j^\dagger \right) + \sum_{ij} \sum_{\mu=x,y,z} \mathrm{Tr} \left(- \sigma^\mu \Psi_i u_{ij}^{\mu *} \Psi_j^\dagger \right)
 \label{timereversaltransformHSLMFmatrix}
\end{aligned}
\end{equation}
Again, with the $SU(2)$ gauge redundancy, that $\mathcal{T}$ being a symmetry requires
\begin{subequations}
\begin{align}
u_{ij}^0 &= G_\mathcal{T} (i) u_{ij}^{0 *} G_\mathcal{T} (j)^\dagger \label{singlettimepsg} \\
u_{ij}^{\mu=x,y,z} &= - G_\mathcal{T} (i) u_{ij}^{\mu *} G_\mathcal{T} (j)^\dagger \label{triplettimepsg}
\end{align}
\end{subequations}
Recall that the collection of the compound operators $G_X X$ (the symmetry group $\lbrace X \rbrace$ now includes both the space group elements and the time reversal) is known as the projective symmetry group (PSG), and we say that the symmetry $X$ is realized projectively if $G_X$ is nontrivial.

In the $\mathbb{Z}_2 \mathrm{U}$ uniform ansatz, for every symmetry $X$ of the system, we set the corresponding $SU(2)$ gauge transformation $G_X=1$ to be trivial. We now investigate how the various symmetries limit the form of the spinon hopping and pairing parameters $\chi_{ij}$, $\Delta_{ij}$, $\mathbf{E}_{ij}$, and $\mathbf{D}_{ij}$. First, time reversal symmetry constrains the singlet parameters $\chi_{ij}$ and $\Delta_{ij}$ to be real, and the triplet parameters $E_{ij}^\mu$ and $D_{ij}^\mu$ to be imaginary, by \eqref{singlettimepsg} and \eqref{triplettimepsg}. We also have $\mu_2=0$. Next, consider the bond $\langle 01 \rangle$, which is mapped to itself under $C_2^x$. By \eqref{singletspacepsg} and \eqref{tripletspacepsg}, we have $u_{10}^0 = u_{01}^0$, $u_{10}^x = u_{01}^x$, $u_{10}^y = -u_{01}^y$, and $u_{10}^z = -u_{01}^z$. As the singlet and triplet parameters obey the relations $\chi_{ji}=\chi_{ij}^*$, $\Delta_{ji}=\Delta_{ij}$, $E_{ji}^\mu=E_{ij}^{\mu *}$, $D_{ji}^\mu=-D_{ij}^\mu$ according to the definitions \eqref{singlethopping}-\eqref{triplethopping}, this implies $E_{01}^x=0$ and $D_{01}^x=0$. The bond $\langle 01 \rangle$ is also mapped to itself under the reflection $\sigma_\mathrm{d}^{[011]}$, which, similar to the analysis of the effect of $C_2^x$ above, constrains $u_{01}^z=-u_{01}^y$, or $E_{01}^z=-E_{01}^y$ and $D_{01}^z=-D_{01}^y$. Finally, we can use $C_3$ or other symmetries to relate the mean field parameters on other bonds to those on $\langle 01 \rangle$. For the singlet parameters it is easy, $\chi_{ij} = \chi_{01}$ and $\Delta_{ij} = \Delta_{01}$ for all bonds $\langle ij \rangle$ by \eqref{singletspacepsg}. For the triplet parameters, we give an example below,
\begin{equation} \label{tripletrelation0102}
\begin{pmatrix} u_{02}^x \\ u_{02}^y \\ u_{02}^z \end{pmatrix} = O_{C_3^{[111]}} \begin{pmatrix} u_{01}^x \\ u_{01}^y \\ u_{01}^z \end{pmatrix} = \begin{pmatrix} u_{01}^z \\ u_{01}^x \\ u_{01}^y \end{pmatrix}
\end{equation}
or $E_{02}^x=-E_{01}^y$, $D_{02}^x=-D_{01}^y$, $E_{02}^y=0$, $D_{02}^y=0$, $E_{02}^z=E_{01}^y$, and $D_{02}^z=D_{01}^y$, by \eqref{tripletspacepsg}. The bond parameters on a down tetrahedron are the same as their counterparts on an up tetrahedron, i.e.~$u_{ij \in \mathrm{down}}^\mu = u_{ij \in \mathrm{up}}^\mu$, by inversion symmetry. There is no further constraint from symmetries, and the number of independent mean field parameters $\chi_{01}$, $\Delta_{01}$, $E_{01}^y$, and $D_{01}^y$ in the $\mathbb{Z}_2 \mathrm{U}$ ansatz is four, as claimed in Section \ref{spinliquidansatzes}.

\subsection{\label{U1Mansatzdetails}$U(1)$ Monopole Flux Ansatz}

The analysis of the $U(1) \mathrm{M}$ ansatz is in some way easier than that of the $\mathbb{Z}_2 \mathrm{U}$ ansatz because the pairing terms $\Delta_{ij}$ and $\mathbf{D}_{ij}$ are zero. There is no need to introduce the matrix \eqref{spinonmatrix} and write down the mean field Hamiltonian in the form \eqref{HSLMFmatrix}. We have instead
\begin{equation} \label{hoppingHSLMF}
H_\mathrm{SL}^\mathrm{MF} = \sum_{ij} \sum_{\mu=0,x,y,z} u_{ij}^\mu \sum_{\alpha \beta} f_{i \alpha}^\dagger \left[ \sigma^\mu \right]_{\alpha \beta} f_{j \beta} + \mathrm{h.c.} ,
\end{equation}
where $u_{ij}^\mu$ are now numbers that depends on the hopping terms instead of matrices. For instance, on the bond $\langle 01 \rangle$, with the exchange couplings $J,K,\Gamma<0$,
\begin{subequations}
\begin{align*}
u_{01}^0 &= - \frac{\lvert \Gamma \rvert}{8} \chi_{01}^* \\
u_{01}^x &= - \frac{\lvert J \rvert}{4} E_{01}^{x *} - \frac{\lvert \Gamma \rvert}{8} E_{01}^{x *} \\
u_{01}^y &= - \frac{\lvert J \rvert}{4} E_{01}^{y *} - \frac{\lvert K \rvert}{8} E_{01}^{y *} - \frac{\lvert \Gamma \rvert}{8} \left( E_{01}^{y *} - E_{01}^{z *} \right) \\
u_{01}^z &= - \frac{\lvert J \rvert}{4} E_{01}^{z *} - \frac{\lvert K \rvert}{8} E_{01}^{z *} - \frac{\lvert \Gamma \rvert}{8} \left( E_{01}^{z *} - E_{01}^{y *} \right)
\end{align*}
\end{subequations}
In the spinon representation of spins \eqref{secondquantizationspinoperator}, for an element $X$ of the space group, the symmetry transformation \eqref{transformspin} is achieved by
\begin{equation} \label{transformspinonoperator}
\begin{pmatrix} f_{i \uparrow} \\ f_{i \downarrow} \end{pmatrix} \overset{X}{\longrightarrow} e^{i \sigma \cdot \hat{\mathbf{n}} \phi / 2} \begin{pmatrix} f_{X(i) \uparrow} \\ f_{X(i) \downarrow} \end{pmatrix} ,
\end{equation}
where $\hat{\mathbf{n}}$ and $\phi$ are as defined previously. However, as mentioned in Section \ref{spinliquidansatzes}, not all of the 48 elements of the space group $\mathrm{Fd}\bar{3}\mathrm{m}$ are respected in the $U(1) \mathrm{M}$ ansatz. The $24$ elements that correspond to inversion, reflections (including glide symmetries), and improper rotations are broken, while the $24$ elements that correspond to proper rotations (including screw symmetries) are realized within the simple PSG constructed in Ref.~\onlinecite{PhysRevB.79.144432}, where the site dependent gauge transformations $G_X = \pm 1$. The proper rotations are,\cite{PhysRevB.79.144432} with the coordinate system defined in Fig.~\ref{cubetetrahedronfigure},
\begin{center}
\begin{tabular}{rp{6.8 cm}}
$e$: & the identity; \\
$8$ $C_3$: & rotation by $\pm 2\pi/3$ about one of the local $[111]$ axes (the directions along the center to the corners of the tetrahedron); \\
$3$ $C_2$: & rotation by $\pi$ about one of the cubic axes ($x$, $y$ and $z$ directions); \\
$6$ $\widetilde{C}_4$: & screw symmetry about one of the axes which are (i) parallel to $x$ axis and going through $(0,a/4,0)$, (ii) parallel to $y$ axis and going through $(0,0,a/4)$, and (iii) parallel to $z$ axis and going through $(a/4,0,0)$ - rotation by $\mp \pi/2$ about one of these axes followed by translation by $a/4$ along that axis; \\
$6$ $\widetilde{C}_2$: & screw symmetry about one of the edges (which connects two sublattices) of a tetrahedron - rotation by $\pi$ about one of the edges followed by translation along that edge.
\end{tabular}
\end{center}

\begin{figure}
\subfloat[]{\label{monopolefluxansatzfigurea} \includegraphics[scale=0.3]{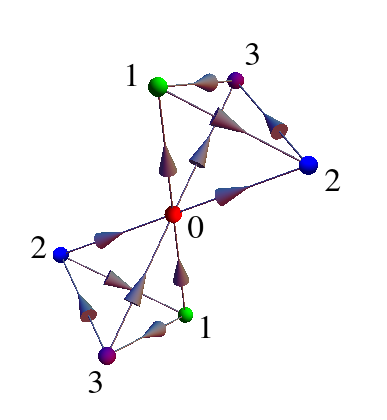}}
\subfloat[]{\label{monopolefluxansatzfigureb} \includegraphics[scale=0.3]{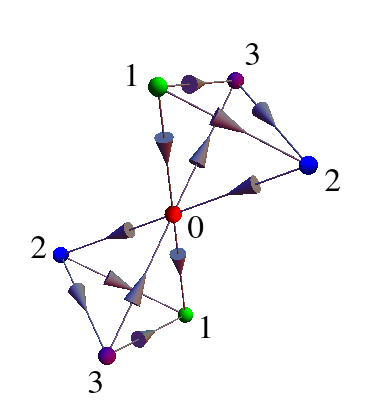}}
\caption{(a) The configuration of link fields $a_{ij}$, which are the arguments of the singlet hopping parameters $\chi_{ij}$ (see \eqref{monopolesinglethopping}), in the monopole flux ansatz. For each link connecting two sites $i$ and $j$, $a_{ij}$ is equal to $\pi/2$ ($-\pi/2$) along (against) the direction of the arrow. This gives a flux of $\pi/2$ on each elementary triangle. (b) The ansatz changes under a symmetry transformation $X$, for example $X=C_2^x$ as shown here. To restore the original configuration of link fields, we apply a sublattice dependent gauge transformation $G_X=\pm 1$, for example $G_{C_2^x} (0) = +1$, $G_{C_2^x} (1) = -1$, $G_{C_2^x} (2) = -1$, and $G_{C_2^x} (3) = +1$. The compound operators $G_X X$, which leave the mean field ansatz invariant, form the monopole flux PSG (see Table \eqref{monopolefluxpsgtable}).}
\end{figure}

It is worth noting that the screw symmetries can be obtained by combining the improper rotations or the reflections with inversion, e.g. $\widetilde{C}_4^x = \mathcal{I} S_4^x$ and $\widetilde{C}_2^{\langle 01 \rangle} = \mathcal{I} \sigma_\mathrm{d}^{[011]}$. The $SO(3)$ matrices corresponding to the $SU(2)$ spin rotations generated by $\widetilde{C}_4^x$ and $\widetilde{C}_2^{\langle 01 \rangle}$ are
\begin{equation} \label{so3matrixscrew}
O_{\widetilde{C}_4^x} = \begin{pmatrix} 1 & 0 & 0 \\ 0 & 0 & 1 \\ 0 & -1 & 0 \end{pmatrix}, \, O_{\widetilde{C}_2^{\langle 01 \rangle}} = \begin{pmatrix} - 1 & 0 & 0 \\ 0 & 0 & 1 \\ 0 & 1 & 0 \end{pmatrix} .
\end{equation}
The monopole flux ansatz is first constructed for the nearest neighbor antiferromagnetic Heisenberg model on the pyrochlore lattice,\cite{PhysRevB.79.144432} where $u_{ij}^0 \sim \lvert J \rvert \chi_{ij}$ and $u_{ij}^{\mu=x,y,z}=0$ in \eqref{hoppingHSLMF}. The singlet hopping parameter between two sites $i$ and $j$ takes the form
\begin{equation} \label{monopolesinglethopping}
\chi_{ij} = \rho e^{i a_{ij}} , \, \rho > 0 , \, a_{ij} = \pm \frac{\pi}{2}.
\end{equation}
The configuration of the link fields $a_{ij}$ is visualized in Fig.~\ref{monopolefluxansatzfigurea}, such that along (against) the direction of the arrow from site $i$ to $j$, $a_{ij}$ equals to $+\pi/2$ ($-\pi/2$). This gives a flux of
\begin{equation} \label{triangleflux}
\sum_{\langle ij \rangle \in \triangle} a_{ij} = \frac{\pi}{2} ,
\end{equation}
on each elementary triangle, if the orientation of the surface is chosen to be pointing towards from the center of the tetrahedron. This can be thought of as a monopole of strength $2 \pi$ sitting inside each tetrahedron, hence the name monopole flux state. Let $X$ be any of the $24$ symmetry elements. Then, for the mean field Hamiltonian of the AFM Heisenberg model,
\begin{equation} \label{transformHAFMMF}
\begin{aligned}[b]
H_\mathrm{HAFM}^\mathrm{MF} & \sim \sum_{ij} f_{i \alpha}^\dagger \chi_{ij} f_{j \alpha} \\
& \overset{X}{\longrightarrow} \sum_{ij} f_{X(i) \alpha}^\dagger \chi_{ij} f_{X(j) \alpha} \\
& \overset{G_X}{\longrightarrow} \sum_{ij} G_X \left( X (i) \right) f_{X(i) \alpha}^\dagger \chi_{ij} f_{X(j) \alpha} G_X \left( X (j) \right),
\end{aligned}
\end{equation}
where the site dependent gauge transformation $G_X=\pm 1$ is introduced to restore the ansatz (i.e.~the original configuration of link fields),
\begin{equation} \label{monopolefluxpsg}
G_X \left( X (i) \right) \chi_{ij} G_X \left( X (j) \right) = \chi_{X(i) X(j)} .
\end{equation}
For example, under $C_3^{[111]}$, the ansatz remains the same as in Fig.~\ref{monopolefluxansatzfigurea}, so $G_{C_3^{[111]}} \left( s \right) = +1$ for all sublattices $s$. However, under $C_2^x$ the configuration of link fields changes to that as in Fig.~\ref{monopolefluxansatzfigureb}, so we choose $G_{C_2^x} \left( 0 \right) = +1$, $G_{C_2^x} \left( 1 \right) = -1$, $G_{C_2^x} \left( 2 \right) = -1$, and $G_{C_2^x} \left( 3 \right) = +1$ to restore the original ansatz. To this end, we summarize the monopole flux PSG $\lbrace G_X X \rbrace$ for all the $24$ symmetry elements $X$ in Table \ref{monopolefluxpsgtable}, similar to Tables $\mathrm{IV}$ and $\mathrm{V}$ in Ref.~\onlinecite{PhysRevB.79.144432}. The monopole flux ansatz is translationally invariant, i.e.~it is the same for every physical unit cell of the pyrochlore lattice.

\begin{table*}
\caption{\label{monopolefluxpsgtable}The projective symmetry group (PSG) of the monopole flux ansatz. The element $G_X X$ is denoted by $X$ for simplicity, where $X$ is one of the $24$ proper rotations (including screw symmetries) of the $\mathrm{Fd\bar{3}m}$ space group. The action of $G_X X$ is shown in \eqref{transformHAFMMF}. The subscript $s$ of the fermionic operator $f_s$ indexes the sublattice.}
\begin{ruledtabular}
\begin{tabular}{cccccccccccc}
$e$ & $C_3^{[111]}$ & $C_3^{[111] \, 2}$ & $C_3^{[1\bar{1}\bar{1}]}$ & $C_3^{[1\bar{1}\bar{1}] \, 2}$ & $C_3^{[\bar{1}1\bar{1}]}$ & $C_3^{[\bar{1}1\bar{1}] \, 2}$ & $C_3^{[\bar{1}\bar{1}1]}$ & $C_3^{[\bar{1}\bar{1}1] \, 2}$ & $C_2^x$ & $C_2^y$ & $C_2^z$  \\ \hline
$f_0$ & $f_0$ & $f_0$ & $f_3$ & $-f_2$ & $f_1$ & $-f_3$ & $f_2$ & $-f_1$ & $-f_1$ & $-f_2$ & $-f_3$ \\
$f_1$ & $f_2$ & $f_3$ & $f_1$ & $f_1$ & $-f_3$ & $f_0$ & $-f_0$ & $-f_2$ & $f_0$ & $-f_3$ & $f_2$ \\
$f_2$ & $f_3$ & $f_1$ & $-f_0$ & $-f_3$ & $f_2$ & $f_2$ & $-f_1$ & $f_0$ & $f_3$ & $f_0$ & $-f_1$ \\
$f_3$ & $f_1$ & $f_2$ & $-f_2$ & $f_0$ & $-f_0$ & $-f_1$ & $f_3$ & $f_3$ & $-f_2$ & $f_1$ & $f_0$ \\ \hline
$\widetilde{C}_4^x$ & $\widetilde{C}_4^{x \, 3}$ & $\widetilde{C}_4^y$ & $\widetilde{C}_4^{y \, 3}$ & $\widetilde{C}_4^z$ & $\widetilde{C}_4^{z \, 3}$ & $\widetilde{C}_2^{\langle 01 \rangle}$ & $\widetilde{C}_2^{\langle 02 \rangle}$ & $\widetilde{C}_2^{\langle 03 \rangle}$ & $\widetilde{C}_2^{\langle 12 \rangle}$ & $\widetilde{C}_2^{\langle 23 \rangle}$ & $\widetilde{C}_2^{\langle 31 \rangle}$ \\ \hline
$f_3$ & $-f_2$ & $f_1$ & $-f_3$ & $f_2$ & $-f_1$ & $f_1$ & $f_2$ & $f_3$ & $-f_0$ & $-f_0$ & $-f_0$ \\
$f_2$ & $f_3$ & $-f_2$ & $-f_0$ & $f_0$ & $-f_3$ & $f_0$ & $f_1$ & $-f_1$ & $f_2$ & $f_1$ & $f_3$ \\
$f_0$ & $-f_1$ & $f_3$ & $f_1$ & $-f_3$ & $-f_0$ & $-f_2$ & $f_0$ & $f_2$ & $f_1$ & $f_3$ & $f_2$ \\
$-f_1$ & $-f_0$ & $f_0$ & $-f_2$ & $f_1$ & $f_2$ & $f_3$ & $-f_3$ & $f_0$ & $f_3$ & $f_1$ & $f_1$ \\
\end{tabular}
\end{ruledtabular}
\end{table*}

Finally, we now extend the monopole flux ansatz to include the triplet hopping parameters, which appears in the mean field Hamiltonian of the nearest neighbor $JK\Gamma$ model on the pyrochlore lattice \eqref{hoppingHSLMF}, using the relation
\begin{equation} \label{tripletspacemonopolepsg}
u_{X(i) X(j)}^\mu = \sum_{\nu} G_X \left( X(i) \right) O_X^{\mu \nu} u_{ij}^\nu G_X \left( X(j) \right) ,
\end{equation}
which can be derived in a way similar to \eqref{transformHSLMFmatrix}. The $SO(3)$ matrices $O_X$ of some representative symmetry elements $X$ can be found in \eqref{so3matrix} and \eqref{so3matrixscrew}.

Since inversion symmetry is broken, the bond parameters of the up and down tetrahedra no longer obey $u_{ij \in \mathrm{up}}^\mu = u_{ij \in \mathrm{down}}^\mu$ as in the $\mathbb{Z}_2 \mathrm{U}$ ansatz. We define $v_{ij}^\mu$ as $u_{ij}^\mu$ for the bond $\langle ij \rangle$ on a down tetrahedron. The form of the singlet hopping parameter $\chi_{ij}$ has already been fixed by \eqref{monopolesinglethopping}. For the triplet hopping parameters on the bond $\langle 01 \rangle$, $C_2^x$ constrains $u_{10}^x = - u_{01}^x$, $u_{10}^y = u_{01}^y$, and $u_{10}^z = u_{01}^z$ by \eqref{tripletspacemonopolepsg}, which implies $E_{01}^x$ is imaginary, while $E_{01}^y$ and $E_{01}^z$ are real. $u_{ij}^\mu$ on other bonds are related to $u_{01}^\mu$ by $C_3$, while $v_{ij}^\mu$ are related to $u_{ij}^\mu$ by $\widetilde{C}_4$ or $\widetilde{C}_2$. The number of independent mean field parameters is four, as claimed in Section \ref{spinliquidansatzes}. There is no further constraint from symmetries. In the absence of pairing channel, for a free fermion hopping Hamiltonian like \eqref{hoppingHSLMF} at zero temperature, the single occupancy constraint is satisfied (on average) by half filling of the momentum states, so there is no need to introduce extra Lagrange multipliers (though $\mu_3$ is often identified with the Fermi level in literature).

\section{\label{localglobalcomparison}Comparisons between the Local and Global Minima from the Mean Field Self Consistent Calculations}

\begin{table}
\caption{\label{Z2Ugaulintable}Comparison between the local and global minima of the mean field Hamiltonian \eqref{HMF} at Gaulin parametrization with the $\mathbb{Z}_2 \mathrm{U}$ ansatz.}
\begin{ruledtabular}
\begin{tabular}{c|ccc|ccc}
 & \multicolumn{3}{c|}{local minimum} & \multicolumn{3}{c}{global minimum} \\ \hline
$B_z/\lvert J_3 \rvert$ & $E$ & phase & $S/S_0$ & $E$ & phase & $S/S_0$ \\ \hline
$0$ & $-0.439$ & $\mathrm{SL}$ & $0$ & $-0.442$ & $\mathrm{FM}^*$ & $0.975$ \\
$0.002$ & $-0.439$ & $\mathrm{FM}^*$ & $0.021$ & $-0.454$ & $\mathrm{FM}^*$ & $0.996$ \\
$0.004$ & $-0.439$ & $\mathrm{FM}^*$ & $0.043$ & $-0.465$ & $\mathrm{FM}$ & $1$ \\
$0.006$ & $-0.440$ & $\mathrm{FM}^*$ & $0.066$ & $-0.477$ & $\mathrm{FM}$ & $1$ \\
$0.008$ & $-0.441$ & $\mathrm{FM}^*$ & $0.091$ & $-0.489$ & $\mathrm{FM}$ & $1$ \\
$0.010$ & $-0.442$ & $\mathrm{FM}^*$ & $0.122$ & $-0.501$ & $\mathrm{FM}$ & $1$ \\
\end{tabular}
\end{ruledtabular}
\end{table}

\begin{table}
\caption{\label{Z2Ucoldeatable}Comparison between the local and global minima of the mean field Hamiltonian \eqref{HMF} at Coldea parametrization with the $\mathbb{Z}_2 \mathrm{U}$ ansatz.}
\begin{ruledtabular}
\begin{tabular}{c|ccc|ccc}
 & \multicolumn{3}{c|}{local minimum} & \multicolumn{3}{c}{global minimum} \\ \hline
$B_z/\lvert J_3 \rvert$ & $E$ & phase & $S/S_0$ & $E$ & phase & $S/S_0$ \\ \hline
$0$ & $-0.471$ & $\mathrm{SL}$ & $0$ & $-0.481$ & $\mathrm{AFM}^*$ & $0.721$ \\
$0.002$ & $-0.471$ & $\mathrm{FM}^*$ & $0.017$ & $-0.481$ & $\mathrm{AFM}^*$ & $0.726$ \\
$0.004$ & $-0.471$ & $\mathrm{FM}^*$ & $0.034$ & $-0.483$ & $\mathrm{M}^*$ & $0.736$ \\
$0.006$ & $-0.472$ & $\mathrm{FM}^*$ & $0.052$ & $-0.485$ & $\mathrm{M}^*$ & $0.756$ \\
$0.008$ & $-0.473$ & $\mathrm{FM}^*$ & $0.071$ & $-0.489$ & $\mathrm{M}^*$ & $0.791$ \\
$0.010$ & $-0.474$ & $\mathrm{FM}^*$ & $0.091$ & $-0.495$ & $\mathrm{FM}^*$ & $0.952$ \\
\end{tabular}
\end{ruledtabular}
\end{table}

We tabulate the energy per unit cell $E$, the phase, and the reduction of magnetic order parameter in magnitude relative to its classical value $S/S_0$ (see the discussion in Sec.~\ref{phasediagram}), of the local and global minima, which correspond to a pure spin liquid/spin liquid dominant and pure magnetic order/magnetic order dominant phases respectively, at various magnetic field strength $B_z$, for the $\mathbb{Z}_2 \mathrm{U}$ ansatz, at Gaulin and Coldea parametrizations (see Tables \ref{Z2Ugaulintable} and \ref{Z2Ucoldeatable}). A representative value $r=0.23$ of the weighting factor is chosen. $S/S_0 \longrightarrow 0$ indicates that the magnetic order is very weak and the system is highly quantum, while $S/S_0 \longrightarrow 1$ indicates that the system approaches the classical limit. In other words, the ratio $S/S_0$ is a good indicator of the quantumness of the system. As $B_z$ increases, the energy difference between the local and global minima grows more significant. Once $B_z$ exceeds $\sim 0.01 \lvert J_3 \rvert$, the spin liquid dominant solution becomes so unfavorable that the self consistent calculations always yield the completely magnetic solution. Similar comparisons are made for the ${U(1) \mathrm{M}}$ ansatz in Tables \ref{U1Mgaulintable} and \ref{U1Mcoldeatable}.

\begin{table}
\caption{\label{U1Mgaulintable}Comparison between the local and global minima of the mean field Hamiltonian \eqref{HMF} at Gaulin parametrization with the $U(1) \mathrm{M}$ ansatz.}
\begin{ruledtabular}
\begin{tabular}{c|ccc|ccc}
 & \multicolumn{3}{c|}{local minimum} & \multicolumn{3}{c}{global minimum} \\ \hline
$B_z/\lvert J_3 \rvert$ & $E$ & phase & $S/S_0$ & $E$ & phase & $S/S_0$ \\ \hline
$0$ & $-0.474$ & $\mathrm{FM}^*$ & $0.008$ & \multicolumn{3}{c}{same as left} \\
$0.002$ & $-0.474$ & $\mathrm{FM}^*$ & $0.039$ & \multicolumn{3}{c}{same as left} \\
$0.004$ & $-0.475$ & $\mathrm{FM}^*$ & $0.072$ & \multicolumn{3}{c}{same as left} \\
$0.006$ & $-0.476$ & $\mathrm{FM}^*$ & $0.110$ & $-0.477$ & $\mathrm{FM}$ & $1$ \\
$0.008$ & $-0.478$ & $\mathrm{FM}^*$ & $0.161$ & $-0.489$ & $\mathrm{FM}$ & $1$ \\
$0.010$ & $-0.480$ & $\mathrm{FM}^*$ & $0.210$ & $-0.501$ & $\mathrm{FM}$ & $1$ \\
\end{tabular}
\end{ruledtabular}
\end{table}

\begin{table}
\caption{\label{U1Mcoldeatable}Comparison between the local and global minima of the mean field Hamiltonian \eqref{HMF} at Coldea parametrization with the $U(1) \mathrm{M}$ ansatz.}
\begin{ruledtabular}
\begin{tabular}{c|ccc|ccc}
 & \multicolumn{3}{c|}{local minimum} & \multicolumn{3}{c}{global minimum} \\ \hline
$B_z/\lvert J_3 \rvert$ & $E$ & phase & $S/S_0$ & $E$ & phase & $S/S_0$ \\ \hline
$0$ & $-0.509$ & $\mathrm{AFM}^*$ & $0.009$ & \multicolumn{3}{c}{same as left} \\
$0.002$ & $-0.509$ & $\mathrm{FM}^*$ & $0.031$ & \multicolumn{3}{c}{same as left} \\
$0.004$ & $-0.510$ & $\mathrm{FM}^*$ & $0.056$ & \multicolumn{3}{c}{same as left} \\
$0.006$ & $-0.511$ & $\mathrm{FM}^*$ & $0.083$ & \multicolumn{3}{c}{same as left} \\
$0.008$ & $-0.512$ & $\mathrm{FM}^*$ & $0.113$ & \multicolumn{3}{c}{same as left} \\
$0.010$ & $-0.514$ & $\mathrm{FM}^*$ & $0.152$ & \multicolumn{3}{c}{same as left} \\
$0.015$ & $-0.520$ & $\mathrm{FM}^*$ & $0.247$ & $-0.525$ & $\mathrm{FM}$ & $1$ \\
\end{tabular}
\end{ruledtabular}
\end{table}

\section{\label{magneticband}Properties of the Spinon Band Structure in the Pure Magnetically Ordered States}
We explain three aspects of the spinon band structure in the pure magnetic phase (where the magnetic order parameters are finite and the spin liquid parameters are zero): (i) flatness (dispersionless), (ii) symmetry about zero energy, and (iii) four fold degeneracy. Recall that the magnetic order parameters are linear combinations of the spin components (see Section \ref{complexfermionmeanfieldtheory}), so that the mean field Hamiltonian \eqref{HMOMF} takes the form
\begin{equation} \label{HMOMFpauli}
\begin{aligned}[b]
H_\mathrm{MO}^\mathrm{MF} &= \sum_{\mathbf{R}} \sum_{s \in \mathbf{R}} \begin{pmatrix} f_{\mathbf{R},s,\uparrow}^\dagger & f_{\mathbf{R},s,\downarrow}^\dagger \end{pmatrix} \left ( c_s^x \sigma^x + c_s^y \sigma^y + c_s^z \sigma^z\right) \begin{pmatrix} f_{\mathbf{R},s,\uparrow} \\ f_{\mathbf{R},s,\downarrow} \end{pmatrix} \\
&= \sum_{\mathbf{k}} \sum_{s=0,1,2,3} \begin{pmatrix} f_{\mathbf{k},s,\uparrow}^\dagger & f_{\mathbf{k},s,\downarrow}^\dagger \end{pmatrix} \left( \sum_{\mu=x,y,z} c_s^\mu \sigma^\mu \right) \begin{pmatrix} f_{\mathbf{k},s,\uparrow} \\ f_{\mathbf{k},s,\downarrow} \end{pmatrix}
\end{aligned}
\end{equation}
with the coefficients $c_s^\mu \in \mathbb{R}$. Since neither spinon hopping nor pairing at two different sites is present, Fourier transform \eqref{fouriertransform} does not introduce any nontrivial phase factor $e^{i \mathbf{k} \cdot (\mathbf{R}_i - \mathbf{R}_j)}$ in the second equality of \eqref{HMOMFpauli}. This explains the flatness of the spinon bands as the energy eigenvalues are independent of the momentum $\mathbf{k}$. Furthermore, written in the basis $( f_{\mathbf{k}, 0, \uparrow}, f_{\mathbf{k}, 0, \downarrow}, \ldots, f_{\mathbf{k}, 3, \uparrow}, f_{\mathbf{k}, 3, \downarrow} )$, the Hamiltonian matrix is an $8 \times 8$ block matrix whose nonzero blocks are the four $2 \times 2$ matrices along the diagonal. Diagonalization yields the energy eigenvalues
\begin{equation} \label{eigenvaluepauli}
\omega_\mathbf{k} = \pm \sqrt{(c_s^{x})^2 + (c_s^{y})^2 + (c_s^{z})^2} .
\end{equation}
The $\pm$ sign means that the spinon bands are symmetric about the zero level. Finally, from \eqref{eigenvaluepauli} we see that the energy eigenvalues depends on the coefficients $c_s^\mu$ only through the second power. We examine the $\mathrm{FM}$ and $\mathrm{AFM}$ order parameters and find that their respective set of coefficients ${c_s^\mu}$ satisfies $c_s^\mu = \pm c_{s'}^\mu$ for different sublattices $s$ and $s'$. This implies the four fold degeneracy.

\bibliography{reference180601}

\end{document}